\documentclass[
aps,prd,
12pt,%10pt
nopreprintnumbers,
showpacs,
eqsecnum,
nofootinbib
]{revtex4-1}

\usepackage{graphicx}
\usepackage{amssymb}

\begin{document}

\title{Accelerating cosmologies in an integrable model with
noncommutative minisuperspace variables}
\author{Nahomi Kan}\email[]{kan@gifu-nct.ac.jp}
\affiliation{National Institute of Technology, Gifu College,
Motosu-shi, Gifu 501-0495, Japan
}
\author{Masashi Kuniyasu}\email[]{mkuni13@yamaguchi-u.ac.jp}
\author{Kiyoshi Shiraishi}\email[]{shiraish@yamaguchi-u.ac.jp}
\author{Kohjiroh Takimoto}\email[]{i016vb@yamaguchi-u.ac.jp}
\affiliation{
Graduate School of Sciences and Technology for Innovation, Yamaguchi
University, Yamaguchi-shi, Yamaguchi 753--8512, Japan}
\date{\today}
%\date{}

\begin{abstract}
We study classical and quantum noncommutative cosmology with
a Liouville-type scalar degree of freedom.
The noncommutativity is imposed on the minisuperspace variables
through a deformation of the Poisson algebra.
In this paper, we investigate the effects of
noncommutativity of minisuperspace variables on the accelerating behavior of 
the cosmic scale factor.
The probability distribution in noncommutative quantum cosmology is also studied
and we propose a novel candidate for interpretation of the probability
distribution in terms of noncommutative arguments.
\end{abstract}

%\preprint{}

\pacs{%
%02.10.Ox, %%%Combinatorics; graph theory
%02.20.Sv, %Lie algebra of Lie groups
%02.30.Hq, %Ordinary differential equations
%02.30.Jr, %Partial differential equations
02.40.Gh, %Noncommutative geometry
%04.20.-q, %%%Classical general relativity
%04.20.Fy, %%Canonical formalism, Lagrangians, and variational principles
04.20.Jb, %%Exact solutions
%04.25.-g, %Approximation
%04.25.Nx, %%%Post-Newtonian approximation; perturbation theory; related
%approximations
%04.40.-b, %Self-Gravitating systems
%04.40.Nr, %%Einstein-Maxwell spacetime
04.50.-h, %%%%%Higher-dimensional gravity and other theories of gravity 
%04.50.Cd, %Kaluza-Klein theories 
%04.50.Gh, %Higher-dimensional black holes, black strings, 
%and related objects 
04.50.Kd, %%%Modified theories of gravity 
%04.60.-m, %%Quantum gravity
04.60.Kz, %%Lower dimensional models; minisuperspace models
%04.60.Rt, %Topologically massive gravity
%04.62.+v, %Quantum fields in curved spacetime
%04.65.+e, %Supergravity
%04.70.Bw, %%%Classical black holes
%05.30.Jp, %Boson systems
%11.10.-z, %%%Field theory
%11.10.Lm, %%%Nonlinear or nonlocal theories and models 
%11.10.Nx, %%%Noncommutative field theory 
%11.10.Kk, %%%Field theories in dimensions other than four
%11.25.-w, %Strings and branes
%11.25.Mj, %%Compactification and four-dimensional models
%11.27.+d %%Extended classical solutions; cosmic strings, 
%domain walls, texture 
%11.30.-j, %Symmetry and conservation laws
%11.30.Pb, %Supersymmetry
%12.60.-i, %Models beyond the standard model
%95.35.+d, %Dark matter
%95.36.+x, %Dark energy
%98.80.-k, %%%Cosmology 
%98.80.Cq, %%%%%Particle-theory and field-theory models of the early
%Universe  
%98.80.Dr, %Relativistic cosmology 
98.80.Qc, %Quantum cosmology
98.80.Jk% %%Mathematical and relativistic aspects of cosmology
.}

\maketitle

%%%%%%%%%%%%%%%%%%%%%%%%%%%%%%%%%%%%%%%%%%%%%%%%%%%%%%%%%%%%%%%%%%%%%%%%%%%
%%%%%%%%%%%%%%%%%%%%%%%%%%%%%%%%%%%%%%%%%%%%%%%%%%%%%%%%%%%%%%%%%%%%%%%%%%%
%%%%%%%%%%%%%%%%%%%%%%%%%%%%%%%%%%%%%%%%%%%%%%%%%%%%%%%%%%%%%%%%%%%%%%%%%%%
\section{Introduction}
\label{sec1}
%%%%%%%%%%%%%%%%%%%%%%%%%%%%%%%%%%%%%%%%%%%%%%%%%%%%%%%%%%%%%%%%%%%%%%%%%%%
%%%%%%%%%%%%%%%%%%%%%%%%%%%%%%%%%%%%%%%%%%%%%%%%%%%%%%%%%%%%%%%%%%%%%%%%%%%
%%%%%%%%%%%%%%%%%%%%%%%%%%%%%%%%%%%%%%%%%%%%%%%%%%%%%%%%%%%%%%%%%%%%%%%%%%%

Almost two decades ago, the noncommutativity of the spacetime coordinates,
such as
\begin{equation}
[x^\mu, x^\nu]=i\theta^{\mu\nu}\,,
\end{equation}
has been introduced into the study of quantum field theory \cite{DN,Szabo}. 
If the parameters of the noncommutativity $\theta$ is taken to be constants,
this means the existence of an absolute small scale unit, $\sim\sqrt{|\theta|}$.
Indeed, motivation of proposing the noncommutative spacetime was originated from
string theory. It is then natural to consider that the noncommutativity may play
some key role in scenarios of quantum gravitation theory.

Since the seminal paper Ref.~\cite{GOR} appeared,
many authors have studied noncommutative quantum
cosmologies~\cite{BP,PM,PO,AAOSS,GSS1,GSS2,GSS3,MOS1,MOS2,OMSS,SPOA,BBDP1,
BBDP2,BBDP3,MMP,OQ}.
In the noncommutative quantum cosmological scenario, 
the authors considered deformation of the minisuperspace variables
instead of deformation of the spacetime algebra, which is a hard task
to treat neatly.
Therefore, it is the simplest way to incorporate noncommutative effects
into a cosmological model as a gravitating system.

On the other hand, 
it is known that the expansion rate of our universe is accelerating in the present
days \cite{darkenergy1,darkenergy2} and also in the very early era of cosmology
\cite{inflation}. Such accelerations can be caused by the dynamics of additional
scalar modes in the Einstein gravity.
In several models, exact solutions are known and thoroughly investigated
\cite{TW,Ohta2,CGG,Roy,Neupane,Russo,MiPi,PiMi,ALNW,KKST}. 

In the present paper, we study classical and quantum noncommutative cosmology
with a Liouville-type scalar degree of freedom.
The exponential scalar potential naturally appears in string 
theory. It is known that such a scalar mode also arises from compactification of
extra dimensions or pure $R^p$ gravity \cite{KSY}. Specifically, we investigate the
effects of noncommutativity in minisuperspace variables on the accelerating
behavior of  the cosmic scale factor in exact analytical solutions of the model.

We suppose that the interpretation of probability in noncommutative quantum
cosmology confronts a subtle problem on variables.
In the commutative case, the Wheeler--De~Witt equation comes from the
Hamiltonian which is represented with the dynamical commutative variables
and derivatives with respect to them. Thus, the arguments of the wave function of
the universe are not the original noncommutative variables.
We should know at least the correspondence between a deformed set of commutative
variables and that of noncommutative variables.
By utilizing the exact wave function in the present model, we consider a
possibility in interpretation of the probability distribution in noncommutative
quantum cosmology.
The possibility relies on the use of the Wigner function \cite{Wigner}.
Since the Wigner function was originally introduced by Wigner in order to
treat the quantum statistical physics appropriately,
it has been widely applied to the problem in various fields of physics and
applied physics \cite{WF}. 

This paper is outlined as follows.
In Sec.~\ref{model}, we define the field-theory action of our toy model and
describe that equivalent actions can be obtained from higher-dimensional theories
and higher-derivative theories.
The action for minisuperspace variables of the model are exhibited in
Sec.~\ref{mini}. 

%v2
We first treat the noncommutative cosmology classically in
Sec.~\ref{class}.
%v2
In Sec.~\ref{clcom}, we review the exact classical commutative
cosmological solutions.
In Sec.~\ref{clnoncom}, we derive the exact classical noncommutative cosmological
solution. We give the deformation of the Poisson algebra here and we find that
the calculations can be performed analytically.
In Sec.~\ref{ac}, we study the effect of the noncommutativity on the accelerating
universe by using the analytical solutions obtained in the previous section.

%v2
Next, we consider the noncommutative quantum cosmology of our model in
Sec.~\ref{quant}.
%v2
In Sec.~\ref{swf}, the wave function of the universe in our noncommutative model is
obtained and connection to the classical solution is described.
Section~\ref{wigner} contains a brief description of the Wigner distribution
function and deformation in minisuperspace variables. 
We propose a new interpretation of the distribution function with respect to
noncommutative variables. 

%v2
Finally, Sec.~\ref{do} is devoted to discussion and
outlook.

%%%%%%%%%%%%%%%%%%%%%%%%%%%%%%%%%%%%%%%%%%%%%%%%%%%%%%%%%%%%%%%%%%%%%%%%%%%
%%%%%%%%%%%%%%%%%%%%%%%%%%%%%%%%%%%%%%%%%%%%%%%%%%%%%%%%%%%%%%%%%%%%%%%%%%%
%%%%%%%%%%%%%%%%%%%%%%%%%%%%%%%%%%%%%%%%%%%%%%%%%%%%%%%%%%%%%%%%%%%%%%%%%%%
\section{The simplest models with exponential scalar potentials}
\label{model}
%%%%%%%%%%%%%%%%%%%%%%%%%%%%%%%%%%%%%%%%%%%%%%%%%%%%%%%%%%%%%%%%%%%%%%%%%%%
%%%%%%%%%%%%%%%%%%%%%%%%%%%%%%%%%%%%%%%%%%%%%%%%%%%%%%%%%%%%%%%%%%%%%%%%%%%
%%%%%%%%%%%%%%%%%%%%%%%%%%%%%%%%%%%%%%%%%%%%%%%%%%%%%%%%%%%%%%%%%%%%%%%%%%%
Let us consider the action of the $D$-dimensional model $(D>2)$%v3
\footnote{We consider here cosmological models derived from the action with the
Einstein--Hilbert term and the Lagrangian for a Liouville-type scalar field. When
we assume $D=2$, the scale factor is not dynamical at least classically.}
\begin{equation}
S=\int d^Dx\sqrt{-g}\left[R-\frac{1}{2}(\nabla\Phi)^2-
\frac{V}{2}e^{2\alpha\Phi}\right]\,,
\label{exp}
\end{equation}
where  $R$ is the Ricci scalar derived from the metric $g_{\mu\nu}$
($\mu,\nu=0,1,\dots,D-1$), $g$ is the determinant of $g_{\mu\nu}$, and $\Phi$ is a
real scalar field. The constant $\alpha$ denotes the scalar self-coupling.
We use the abbreviation
$(\nabla\Phi)^2\equiv g^{\mu\nu}\partial_\mu\Phi\partial_\nu\Phi$.
%R
Further, we assume that $V$ is a constant.
%R 
It is known that the exponential potential arises from string theory and is found
in string-motivated field theory \cite{PW,Mignemi}%
\footnote{It is also worth noting that exponential-type potentials appear in
general dynamical models of cosmological evolution \cite{Simeone}.}. In what
follows, we exhibit several equivalent models.
%%%%%%%%%%%%%%%%%%%%%%%%%%%%%%%%%%%%%%%%%%%%%%%%%%%%%%%%%%%%%%%%%%%%%%%%%%%
%%%%%%%%%%%%%%%%%%%%%%%%%%%%%%%%%%%%%%%%%%%%%%%%%%%%%%%%%%%%%%%%%%%%%%%%%%%
\subsection{Higher-dimensional Einstein-antisymmetric field theory with a
cosmological constant}
%%%%%%%%%%%%%%%%%%%%%%%%%%%%%%%%%%%%%%%%%%%%%%%%%%%%%%%%%%%%%%%%%%%%%%%%%%%
%%%%%%%%%%%%%%%%%%%%%%%%%%%%%%%%%%%%%%%%%%%%%%%%%%%%%%%%%%%%%%%%%%%%%%%%%%%

Let us consider an Einstein-antisymmetric field theory with a cosmological constant
in
$(D+q)$ dimensions. Its well-known action is
\begin{equation}
S_{D+M}=\int d^{D+q}x \sqrt{-g_{(D+q)}}\left[R_{(D+q)}
-\frac{1}{2q!}F_{[q]}^2-2\Lambda\right]\,,
\end{equation}
where ${g_{(D+q)}}_{MN}$ is the determinant of the metric ${g_{(D+q)}}_{MN}$
($M, N=0,\dots,D+q-1$) and $R_{(D+q)}$ is the Ricci scalar constructed from
${g_{(D+q)}}_{MN}$. The constant $\Lambda$ represents the cosmological constant.
The square of $q$-form field strength $F_{[q]}^2$ means 
$g^{M_1 N_1}g^{M_2 N_2}\cdots g^{M_q N_q}$ $
{F_{[q]}}_{M_1M_2\cdots M_q}{F_{[q]}}_{N_1N_2\cdots N_q}$.%
\footnote{Upon compactification, it is also possible to consider $(D-1)$-form
field strength. We omit this possibility in this paper only to consider the
simplest model (and leave the possibility for future study).} 

We take a representation for the
$(D+M)$-dimensional metric such as
\begin{equation}
ds^2={g_{(D+q)}}_{MN}dx^Mdx^N=e^{-\frac{2qb}{D-2}}{g}_{\mu\nu}dx^\mu dx^\nu+
e^{2b}\tilde{g}_{mn}dx^mdx^n\,,
\end{equation}
where $\mu, \nu=0,\dots,D-1$ and $m, n=D,\dots, D+q-1$.
The Ricci tensor of the maximal symmetric extra space, whose metric is
denoted as
$\tilde{g}_{mn}$, is assumed to be written as
\begin{equation}
\tilde{R}_{mn}=k_b(q-1)\tilde{g}_{mn}\,,
\end{equation}
where $k_b$ is a constant, which has been normalized to $1$, $0$, or $-1$.
Further, we assume that the $q$-form field strength takes a constant value in the
extra space; thus, 
\begin{equation}
{F_{[q]}}_{D,D+1,\dots,D+q-1}=f\,,
\end{equation}
where $f$ is a constant.

Performing the dimensional reduction, we find that the effective $D$-dimensional
action is
\begin{eqnarray}
S_{D}&=&\int d^{D}x \sqrt{-g}\left[R-\frac{q(D+q-2)}{D-2}(\nabla
b)^2+q(q-1)k_be^{-2\frac{D+q-2}{D-2}b}\right.
\nonumber \\
& &
\qquad\qquad\qquad\left.
-\frac{f^2}{2}e^{-2\frac{q(D-1)}{D-2}b}
-2\Lambda e^{-2\frac{q}{D-2}b}\right]\,,
\end{eqnarray}
where we omit the overall constant.
Further, defining
\begin{equation}
\Phi=-\sqrt{\frac{2q(D+q-2)}{D-2}}b\,,
\end{equation}
we obtain the effective action of gravitating scalar field with an exponential
potential (\ref{exp}). Then, we identify the coupling parameters
in the following three cases:
\begin{itemize}
\item Case of compactification, only with the curvature term (Case [CC],
$(q-1)k_b\ne 0$ and
$f=0$ and
$\Lambda=0$):
\begin{equation}
\alpha=\sqrt{\frac{D+q-2}{2q(D-2)}}\,,\quad
V=-2q(q-1)k_b\,.
\end{equation}
\item Case of compactification, only with the flux term (Case [CF], $f\ne 0$ and
$\Lambda=0$ and $(q-1)k_b=0$):
\begin{equation}
\alpha=(D-1)\sqrt{\frac{q}{2(D-2)(D+q-2)}}\,,\quad
V=f^2\,.
\end{equation}
\item Case of compactification, only with the cosmological (Lambda) term (Case
[CL], $\Lambda\ne 0$ and
$(q-1)k_b=0$ and $f=0$):
\begin{equation}
\alpha=\sqrt{\frac{q}{2(D-2)(D+q-2)}}\,,\quad
V=4\Lambda\,.
\end{equation}
\end{itemize}
%%%%%%%%%%%%%%%%%%%%%%%%%%%%%%%%%%%%%%%%%%%%%%%%%%%%%%%%%%%%%%%%%%%%%%%%%%%
%%%%%%%%%%%%%%%%%%%%%%%%%%%%%%%%%%%%%%%%%%%%%%%%%%%%%%%%%%%%%%%%%%%%%%%%%%%
\subsection{Pure $R^p$ gravity}
%%%%%%%%%%%%%%%%%%%%%%%%%%%%%%%%%%%%%%%%%%%%%%%%%%%%%%%%%%%%%%%%%%%%%%%%%%%
%%%%%%%%%%%%%%%%%%%%%%%%%%%%%%%%%%%%%%%%%%%%%%%%%%%%%%%%%%%%%%%%%%%%%%%%%%%
Now, we turn to consider pure $R^p$ gravity in $D$ dimensional
spacetime \cite{KSY}. We start with the action
\begin{equation}
S_{}=\frac{1}{p}\int
d^Dx\,\sqrt{-g}\,R^p
\,.
\end{equation}
We can use an auxiliary field $\chi$ to obtain
classically equivalent action:
\begin{equation}
S=\int
d^Dx\sqrt{-g}\,\left[{\chi^{p-1}}R-\frac{p-1}{p}\chi^p\right]\,.
\label{rp}
\end{equation}
We can eliminate the $\chi$-dependence in front of the Einstein--Hilbert term $R$
in the action (\ref{rp}) by a Weyl transformation.
In other words, we consider a Weyl-transformed metric $\tilde{g}_{\mu\nu}$
which satisfies
$\sqrt{-g}\chi^{p-1} R=\sqrt{-\tilde{g}}\tilde{R}+\cdots$,
where $\tilde{R}$ is the Ricci scalar constructed from $\tilde{g}_{\mu\nu}$.
To this end, we choose
$g_{\mu\nu}=\chi^{-\frac{2(p-1)}{D-2}}\tilde{g}_{\mu\nu}$ in this time.
Then, we obtain
\begin{equation}
S=\int d^Dx\sqrt{-\tilde{g}}\left[
\tilde{R}-
\frac{1}{2}(\tilde{\nabla}\Phi)^2-\frac{p-1}{p}e^{\sqrt{\frac{2}{(D-1)(D-2)}}
\frac{p-D/2}{p-1}\Phi}\right]\,.
\label{Rp}
\end{equation}
Here, we defined 
$\Phi\equiv-\sqrt{\frac{2(D-1)}{D-2}}(p-1)\ln\chi$.
This action (\ref{Rp}) is equivalent to the action (\ref{exp})
with the coupling constants
\begin{equation}
\alpha=\frac{1}{\sqrt{2(D-1)(D-2)}}
\frac{p-D/2}{p-1}\,,\quad
V=\frac{2(p-1)}{p}\,.
\end{equation}
We call this case as Case [RP], for later convenience.

%%%%%%%%%%%%%%%%%%%%%%%%%%%%%%%%%%%%%%%%%%%%%%%%%%%%%%%%%%%%%%%%%%%%%%%%%%%
%%%%%%%%%%%%%%%%%%%%%%%%%%%%%%%%%%%%%%%%%%%%%%%%%%%%%%%%%%%%%%%%%%%%%%%%%%%
%%%%%%%%%%%%%%%%%%%%%%%%%%%%%%%%%%%%%%%%%%%%%%%%%%%%%%%%%%%%%%%%%%%%%%%%%%%
\section{action for minisuperspace variables}
\label{mini}
%%%%%%%%%%%%%%%%%%%%%%%%%%%%%%%%%%%%%%%%%%%%%%%%%%%%%%%%%%%%%%%%%%%%%%%%%%%
%%%%%%%%%%%%%%%%%%%%%%%%%%%%%%%%%%%%%%%%%%%%%%%%%%%%%%%%%%%%%%%%%%%%%%%%%%%
%%%%%%%%%%%%%%%%%%%%%%%%%%%%%%%%%%%%%%%%%%%%%%%%%%%%%%%%%%%%%%%%%%%%%%%%%%%
Now, we concentrate ourselves on studying cosmological behavior of the model
described by the action (\ref{exp}).
We adopt the following ans\"atze.
The $(D-1)$-dimensional space is assumed to be a flat Euclidean space%
\footnote{If the space has curvature, the effective potential becomes a
complicated one and the simple separation of variables as follows does not occur
in general.}  and its scale factors and the scalar
$\Phi$ are considered to be only time-dependent; i.e., they are functions of the
time coordinate $t=x^0$. Therefore, we take the metric as follows:
\begin{equation}
ds^2=-e^{2n(t)}dt^2+e^{2a(t)}d\mathbf{x}^2\,,
\end{equation}
where $d\mathbf{x}^2\equiv\sum_{\mu=1}^{D-1}(dx^{\mu})^2$.

Substituting the anz\"atze and noting that $\sqrt{-g}\propto
e^{(D-1)a+n}$, we find
\begin{equation}
S\propto
\int dt\,\, e^{(D-1)a-n}\left[2 (D-1)
\ddot{a}+D(D-1)\dot{a}^2-2(D-1)\dot{n}\dot{a} +\frac{1}{2}\dot{\Phi}^2-
\frac{V}{2}e^{2\alpha\Phi}e^{2n}\right]\,,
\end{equation}
where the dot indicates the derivative with respect to time $t$.
Here, if we set
\begin{equation}
n(t)=(D-1) a(t)\,
\end{equation}
as a gauge choice, the reduced cosmological action becomes%
\footnote{It is known that the Gibbons--Hawking--York boundary terms
\cite{York,GH}, which is not explicitly denoted as usual, makes the action
canonical.}
\begin{equation}
S\propto
\int dt
\left[-(D-1)(D-2)\dot{a}^2+\frac{1}{2}\dot{\Phi}^2
-\frac{V}{2} e^{2[(D-1)a+\alpha\Phi]}
\right]\,.
\end{equation}
We now find that the ``kinetic'' terms, which contain the time derivatives, in the
above action can have a form
\begin{eqnarray}
& &
-(D-1)(D-2)\dot{a}^2
+\frac{1}{2}\dot{\Phi}^2\nonumber
\\
&=&-\frac{1}{1-{2\alpha^2}
\frac{D-2}{D-1}}
\left[(D-1)(D-2)\left(\dot{a}+\frac{\alpha}{D-1}\dot{\Phi}\right)^2
-\frac{1}{2}\left(\dot{\Phi}+2(D-2){\alpha}\dot{a}\right)^2\right]
\,.
\end{eqnarray}
Therefore, we can consider the following reduced Lagrangian of two
dynamical variables to analyze the dynamics in minisuperspace: 
\begin{equation}
L=-\frac{1}{2}\dot{x}^2+\frac{1}{2}\dot{y}^2-\frac{U}{2}e^{2\lambda x}\,,
\end{equation}
where
\begin{equation}
x(t)\equiv\sqrt{2(D-1)(D-2)}\left({a}+\frac{\alpha}{D-1}{\Phi}\right)\,,
\quad
y(t)\equiv{\Phi}+2(D-2){\alpha}{a}\,,
\label{xy}
\end{equation}
and
\begin{equation}
\lambda\equiv\sqrt{\frac{D-1}{2(D-2)}}\,,
\quad
U\equiv\Sigma \cdot V\,,
\end{equation}
where
\begin{equation}
\Sigma\equiv1-{2\alpha^2}\frac{D-2}{D-1}\,.
\end{equation}

The canonical conjugate momenta are
\begin{equation}
\pi_x\equiv\frac{\partial L}{\partial\dot{x}}=-\dot{x}\,,\quad
\pi_y\equiv\frac{\partial L}{\partial\dot{y}}=\dot{y}\,,
\end{equation}
and then, the Hamiltonian of the system is found to be
\begin{equation}
H=-\frac{1}{2}\pi_x^2+\frac{1}{2}\pi_y^2+\frac{U}{2}e^{2\lambda x}\,.
\label{Ham}
\end{equation}

The correspondence to the model cases in the previous section is as follows:
\begin{eqnarray}
& &\bullet~\mbox{Case [CC]:}\qquad
U=-2\frac{(D-2)(q-1)^2}{D-1}k_b\,.\\
& &\bullet~\mbox{Case [CF]:}\qquad
U=-\frac{(D-2)(q-1)}{D+q-2}f^2\,.\\
& &\bullet~\mbox{Case [CL]:}\qquad
U=4\frac{(D-2)(D+q-1)}{(D-1)(D+q-2)}\Lambda\,.\\
& &\bullet~\mbox{Case [RP]:}\qquad
U=\frac{(D-2)(2p-1)(2Dp-3D+2)}{2(D-1)^2p(p-1)}\,.\nonumber\\
& &\hspace{3cm}\mbox{(for }D=4, U>0 \mbox{ if }1/2<p<1\mbox{ or }p>5/4.)
\end{eqnarray}

%v2
In the next section, we will give analytical classical solutions for
both commutative and noncommutative system and investigate the cosmic acceleration
in noncommutative classical cosmology.
%v2

%%%%%%%%%%%%%%%%%%%%%%%%%%%%%%%%%%%%%%%%%%%%%%%%%%%%%%%%%%%%%%%%%%%%%%%%%%%
%%%%%%%%%%%%%%%%%%%%%%%%%%%%%%%%%%%%%%%%%%%%%%%%%%%%%%%%%%%%%%%%%%%%%%%%%%%
%%%%%%%%%%%%%%%%%%%%%%%%%%%%%%%%%%%%%%%%%%%%%%%%%%%%%%%%%%%%%%%%%%%%%%%%%%%
\section{The classical system}
\label{class}
%%%%%%%%%%%%%%%%%%%%%%%%%%%%%%%%%%%%%%%%%%%%%%%%%%%%%%%%%%%%%%%%%%%%%%%%%%%
%%%%%%%%%%%%%%%%%%%%%%%%%%%%%%%%%%%%%%%%%%%%%%%%%%%%%%%%%%%%%%%%%%%%%%%%%%%
%%%%%%%%%%%%%%%%%%%%%%%%%%%%%%%%%%%%%%%%%%%%%%%%%%%%%%%%%%%%%%%%%%%%%%%%%%%

%%%%%%%%%%%%%%%%%%%%%%%%%%%%%%%%%%%%%%%%%%%%%%%%%%%%%%%%%%%%%%%%%%%%%%%%%%%
%%%%%%%%%%%%%%%%%%%%%%%%%%%%%%%%%%%%%%%%%%%%%%%%%%%%%%%%%%%%%%%%%%%%%%%%%%%
\subsection{Classical commutative solution}
\label{clcom}
%%%%%%%%%%%%%%%%%%%%%%%%%%%%%%%%%%%%%%%%%%%%%%%%%%%%%%%%%%%%%%%%%%%%%%%%%%%
%%%%%%%%%%%%%%%%%%%%%%%%%%%%%%%%%%%%%%%%%%%%%%%%%%%%%%%%%%%%%%%%%%%%%%%%%%%
%v2
First of all, we review the derivation of commutative classical solutions in the
system. 
%v2
As usual dynamical systems, we will work with
the Poisson brackets 
\begin{equation}
\{x, \pi_x\}=\{y, \pi_y\}=1\,,
\label{cpb}
\end{equation} 
and others are zero.
The usual Hamilton's equations for the Hamiltonian (\ref{Ham}) are
\begin{eqnarray}
& &\dot{x}=\{x, H\}=\frac{\partial H}{\partial \pi_x}=-\pi_x\,,\quad
\dot{y}=\{y, H\}=\frac{\partial H}{\partial \pi_y}=\pi_y\,,\nonumber \\
& &\dot{\pi}_x=\{\pi_x, H\}=-\frac{\partial H}{\partial x}=-\lambda U e^{2\lambda
x}\,,\quad
\dot{\pi}_y=\{\pi_y, H\}=-\frac{\partial H}{\partial y}=0\,.
\end{eqnarray}
From these equations, we obtain the equations of motion as follows:
\begin{equation}
\ddot{x}-\lambda U
e^{2\lambda x}=0\,,\qquad
\label{xeq}
\ddot{y}=0\,.
\label{yeq}
\end{equation}
Note that, because we now consider the
cosmological system, the parametric invariance of $t\rightarrow ct$ ($c$ is a
constant) requires
\begin{equation}
H=0\,,
\label{H0}
\end{equation}
if the solution is substituted into the Hamiltonian.

One can easily find that the solution for $y(t)$ is simply given by
\begin{equation}
y(t)=P (t-t_0)+y_0\,,
\end{equation}
where $P$, $t_0$ and $y_0$ are constants.
The exact solution for $x(t)$, which obeys one-dimensional Liouville equation
and the constraint $H=0$, is given by
\begin{eqnarray}
& &\bullet~U>0\qquad
x(t)=\frac{1}{2\lambda}\ln
\frac{P^2}{U\sinh^2\lambda P (t-t_0)}\,,\\
& &\bullet~U<0\qquad
x(t)=\frac{1}{2\lambda}\ln
\frac{P^2}{|U|\cosh^2\lambda P (t-t_0)}\,.
\end{eqnarray}

%%%%%%%%%%%%%%%%%%%%%%%%%%%%%%%%%%%%%%%%%%%%%%%%%%%%%%%%%%%%%%%%%%%%%%%%%%%
%%%%%%%%%%%%%%%%%%%%%%%%%%%%%%%%%%%%%%%%%%%%%%%%%%%%%%%%%%%%%%%%%%%%%%%%%%%
\subsection{Classical noncommutative system}
\label{clnoncom}
%%%%%%%%%%%%%%%%%%%%%%%%%%%%%%%%%%%%%%%%%%%%%%%%%%%%%%%%%%%%%%%%%%%%%%%%%%%
%%%%%%%%%%%%%%%%%%%%%%%%%%%%%%%%%%%%%%%%%%%%%%%%%%%%%%%%%%%%%%%%%%%%%%%%%%%
%v2
Now, we investigate the noncommutative classical dynamics of the system.
%v2
At first, we define the minisuperspace variables with noncommutativity.
By replacing
\begin{equation}
x\rightarrow X\,,\quad
\pi_x\rightarrow \Pi_X\,,\quad
y\rightarrow Y\,,\quad
\pi_y\rightarrow \Pi_Y\,,
\end{equation}
we obtain the Hamiltonian
\begin{equation}
H_\theta=-\frac{1}{2}\Pi_X^2+\frac{1}{2}\Pi_Y^2+\frac{U}{2}e^{2\lambda X}\,.
\label{ht}
\end{equation}

Here, we consider the Poisson brackets
\begin{equation}
\{X, \Pi_X\}=1\,,\quad
\{Y, \Pi_Y\}=1\,,\quad
\{X, Y\}=\theta\,,
\label{ncpb}
\end{equation}
and others are zero. The parameter of the noncommutativity $\theta$ is assumed to
be a constant. If we set the noncommutativity parameter
$\theta\rightarrow 0$, the system returns to the commutative system.
The Hamilton's equations are now given by
\begin{eqnarray}
& &\dot{X}=\{X, H_\theta\}=-\Pi_X\,,\quad
\dot{Y}=\{Y, H_\theta\}=\Pi_Y-\theta\lambda U e^{2\lambda X} \,,\nonumber\\
& &\dot{\Pi}_X=\{\Pi_X, H_\theta\}=-\lambda Ue^{2\lambda
X}\,,\quad
\dot{\Pi}_Y=\{\Pi_Y, H_\theta\}=0\,.
\label{nche}
\end{eqnarray}

The solution for these equations turns out to be
\cite{BP,PO,GSS1,OMSS,SPOA,BBDP2,BBDP3}
\begin{eqnarray}
& &\bullet ~U>0 \nonumber \\
& &X(t)=\frac{1}{2\lambda}\ln
\frac{P^2}{U\sinh^2\lambda P (t-t_0)}\,,\quad Y(t)=P(t-t_0)+y_0+\theta
P\coth\lambda P(t-t_0)\,,
\label{csp} \\
& &\bullet ~U<0 \nonumber \\
& &X(t)=\frac{1}{2\lambda}\ln
\frac{P^2}{|U|\cosh^2\lambda P (t-t_0)}\,,\quad Y(t)=P(t-t_0)+y_0+\theta
P\tanh\lambda P(t-t_0)\,,
\label{csm}
\end{eqnarray}
which satisfy the Hamiltonian constraint $H_\theta=0$.
It should be noticed that the analytic solutions can be simply classified
by the sign of $U$.

Using the variables $x, y, \pi_x,$ and $\pi_y$ which satisfy
commutative Poisson brackets (\ref{cpb}), we can express the noncommutative
Poisson algebra (\ref{ncpb})
\cite{GOR,BP,PM,PO,AAOSS,GSS1,GSS2,GSS3,MOS1,MOS2,OMSS,SPOA,BBDP1,
BBDP2,BBDP3,MMP,OQ,Djemai}. Here, we develop more generic deformation of
variables. We define
\begin{equation}
X=x-\frac{\theta-\rho}{2}\pi_y\,,\quad Y=y+\frac{\theta+\rho}{2}\pi_x\,,\quad
\Pi_X=\pi_x\,,\quad \Pi_Y=\pi_y\,,
\label{xX}
\end{equation}
where $\rho$ is a constant.
One can easily find that these satisfy noncommutative Poisson brackets
(\ref{ncpb}). Although the symmetric choice $\rho=0$ seems to be natural
especially for the case of the spacetime noncommutativity, there is no reason to
specify the choice in the present small system of minisuperspace.

Now, the Hamiltonian (\ref{ht}) reads
\begin{equation}
H_\theta=-\frac{1}{2}\pi_x^2+\frac{1}{2}\pi_y^2+\frac{U}{2}e^{2\lambda
[x-(\theta-\rho)\pi_y/2]}\,.
\end{equation}
The Hamilton's equations for commutative variables become
\begin{eqnarray}
& &\dot{x}=\frac{\partial H_\theta}{\partial \pi_x}=-\pi_x\,,\quad
\dot{y}=\frac{\partial H_\theta}{\partial \pi_y}=\pi_y
-\frac{\theta-\rho}{2}\lambda U  e^{2\lambda
[x-(\theta-\rho)\pi_y/2]}\,,\nonumber \\
& &\dot{\pi}_x=-\frac{\partial H_\theta}{\partial x}=-\lambda Ue^{2\lambda
[x-(\theta-\rho)\pi_y/2]}\,,\quad
\dot{\pi}_y=-\frac{\partial H_\theta}{\partial y}=0\,,
\end{eqnarray}
Then, the Hamilton's equations (\ref{nche}) are recovered as follows:
\begin{eqnarray}
\dot{X}&=&\dot{x}-\frac{\theta-\rho}{2}\dot{\pi}_y=-\pi_x=-\Pi_X\,,\nonumber\\
\dot{Y}&=&\dot{y}+\frac{\theta+\rho}{2}\dot{\pi}_x=\pi_y
-\theta\lambda U  e^{2\lambda
[x-(\theta-\rho)\pi_y/2]}=\Pi_Y
-\theta\lambda U  e^{2\lambda
X}\,,\nonumber\\
\dot{\Pi}_X&=&\dot{\pi}_x=-\lambda Ue^{2\lambda
X}\,,\quad
\dot{\Pi}_Y=\dot{\pi}_y=0\,.
\end{eqnarray}
Note that there remains no dependence on the value of $\rho$.

Incidentally, the solutions for commuting variables $x$ and $y$ are given by
\begin{eqnarray}
\bullet ~U>0 \quad
& &x(t)=\frac{1}{2\lambda}\ln
\frac{P^2}{U\sinh^2\lambda P (t-t_0)}+\frac{\theta-\rho}{2}P\,,\nonumber \\
& &y(t)=P(t-t_0)+y_0+\frac{\theta-\rho}{2} P\coth\lambda P(t-t_0)\,,
\label{s1}\\
\bullet ~U<0 \quad 
& &x(t)=\frac{1}{2\lambda}\ln
\frac{P^2}{|U|\cosh^2\lambda P (t-t_0)}+\frac{\theta-\rho}{2}P\,,\nonumber \\
& &y(t)=P(t-t_0)+y_0+\frac{\theta-\rho}{2} P\tanh\lambda P(t-t_0)\,,
\label{s2}
\end{eqnarray}
which satisfy $H=0$.
Note that $x(t)$ and $y(t)$ depend only on the combination of the parameters
$\theta-\rho$. Of course, the variables
$X(t)$ and $Y(t)$ constructed from the solutions for $x(t)$ and $y(t)$ with the
relation (\ref{xX}) are independent of the additional parameter $\rho$.

%%%%%%%%%%%%%%%%%%%%%%%%%%%%%%%%%%%%%%%%%%%%%%%%%%%%%%%%%%%%%%%%%%%%%%%%%%%
%%%%%%%%%%%%%%%%%%%%%%%%%%%%%%%%%%%%%%%%%%%%%%%%%%%%%%%%%%%%%%%%%%%%%%%%%%%
\subsection{Effect of noncommutativity on accelerating universe}
\label{ac}
%%%%%%%%%%%%%%%%%%%%%%%%%%%%%%%%%%%%%%%%%%%%%%%%%%%%%%%%%%%%%%%%%%%%%%%%%%%
%%%%%%%%%%%%%%%%%%%%%%%%%%%%%%%%%%%%%%%%%%%%%%%%%%%%%%%%%%%%%%%%%%%%%%%%%%%
%v2
Here, we study the effect of noncommutativity on the accelerating
universe.
%v2
The solutions 
%v2
(\ref{s1}) and (\ref{s2})
%v2
represent the expanding universe
in appropriate ranges of parameters.
To investigate the evolution of the scale factor, we should regard the following
form for the metric:
\begin{eqnarray}
ds^2&=&-d\eta^2+S^2(\eta)d\mathbf{x}^2\,,
\end{eqnarray}
where $\eta$ is the cosmic time for the $D$-dimensional spacetime
and $S$ is the ``physical'' scale factor of $(D-1)$-dimensional flat space. Thus,
we obtain the relations
\begin{equation}
S(\eta)=e^{a(t)}\,,\quad
d\eta=\pm e^{(D-1)a(t)}dt=\pm S^{D-1} dt\,.
\end{equation}
Note that, because of the time-reversal symmetry, we can choose the sign in the
above equation as to observe an expanding phase.

It is difficult to determine the existence or absence of transient acceleration
only from analytic methods. Therefore, we should investigate the behavior of
$S(\eta)$ in a numerical plot. Anyway, analytic solutions are very useful to
express numerical values. To this end, we first observe
\begin{equation}
\frac{dS}{d\eta}=S^{1-D}\frac{dS}{dt}=
-\frac{1}{D-2}\frac{dS^{2-D}}{dt}\,,\quad
\frac{d^2S}{d\eta^2}=-\frac{1}{D-2}S^{1-D}\frac{d^2S^{2-D}}{dt^2}\,.
\end{equation}
Thus, for expanding 
%v2
($\frac{dS}{d\eta}>0$)
%v2
and accelerating physical universes 
%v2
($\frac{d^2S}{d\eta^2}>0$) \cite{Ohta2},
%v2
$-\frac{dS^{2-D}}{dt}>0$ and $-\frac{d^2S^{2-D}}{dt^2}>0$.
%v2
All the solutions in the present model have an expanding phase with an appropriate
choice of the direction of time. We define a normalized function
$A(t)\equiv-S^{D-2}(t)\frac{d^2S^{2-D}(t)}{dt^2}$, which is independent of the
absolute scale of $S$, to investigate the cosmic acceleration \cite{KKST}. Then,
the positive value of $A(t)$ indicates the accelerating phase of the universe.
%v2

In our present parametrization, we find%v3
\footnote{Note that the relation (\ref{xy}) is assumed even for the new variables
$x\rightarrow X$ and $y\rightarrow Y$.
Note also that we only use the solutions obtained above in the following
expressions and no longer worry about the noncommutativity, which affected the
dynamics of variables.}
\begin{equation}
S^{D-2}=e^{(D-2)a}=\exp\left[{\frac{D-2}{(D-1)\Sigma}(\lambda X-\alpha
Y)}\right]\,.
\end{equation}

%%%%%%%%%%%%%%%%%%%%%%%%%%%
% 1
%%%%%%%%%%%%%%%%%%%%%%%%%%%
%\begin{wrapfigure}{r}{5cm}
\begin{figure}[ht]
\centering
\includegraphics[width=8cm]{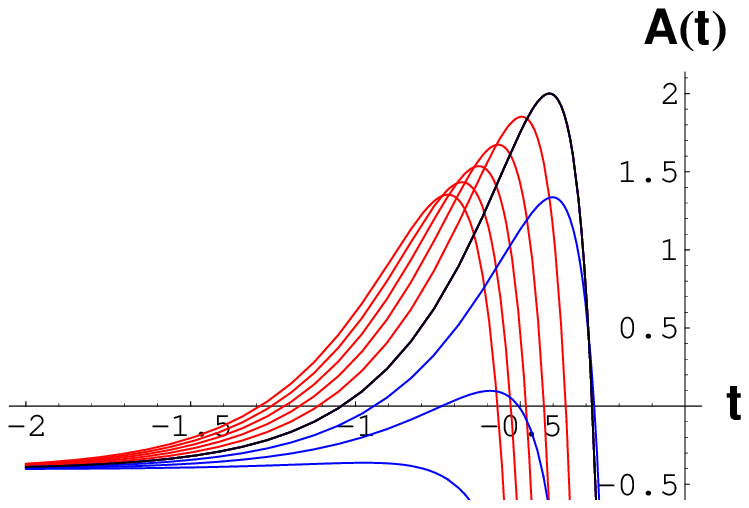}\quad
\includegraphics[width=8cm]{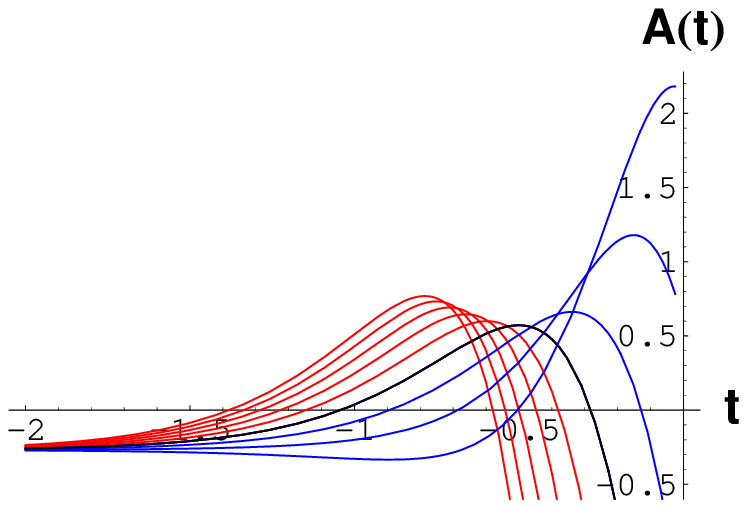}
\\
(a) \hspace{8cm} (b)
\caption{(a) $A(t)$  as a function of $t$ in Case [CC] with $k_b=-1$. 
The curve in black indicates the commutative case ($\theta=0$). The red curves
correspond to the cases with
$\theta=0.1, 0.2, 0.3, 0.4, 0.5$, according to the location of the peak from top
to bottom.  The blue curves
correspond to the cases with
$\theta=-0.1, -0.2, -0.3$, according to the location of the peak from
top to bottom. (b) $A(t)$ as a function of $t$ in Case [CF]. The red curves
correspond to the cases with
$\theta=0.1, 0.2, 0.3, 0.4, 0.5$, according to the location of the peak from bottom
to top.  The blue curves
correspond to the cases with
$\theta=-0.1, -0.2, -0.3$, according to the location of the peak from
bottom to top. For the other parameters for each fugure, see text.}
\label{fig1}
\end{figure}
%\end{wrapfigure}
%%%%%%%%%%%%%%%%%%%%%%%%%%%

We first consider Case [CC], with $k_b=-1$, $D=4$, $q=2$, $t_0=0$
and $P=2$. As already known, the model with $k_b=-1$, i.e., with the hyperbolic
internal space, yields an accelerating universe \cite{TW,Ohta2}. 
We show the values $A(t)=-S^{D-2}(t)\frac{d^2S^{2-D}(t)}{dt^2}$
versus $t$ for $\theta=-0.3\sim 0.5$ in FIG.~\ref{fig1}~(a).
The plot shows that acceleration is weakened for nonzero values of the
noncommutativity parameter $\theta$. For $\theta>0$, we find that the acceleration
period remains for a broad range of parameters. On the other hand, for $\theta<0$,
the acceleration period disappears for sufficient large values of
$|\theta|$. The same tendencies are found in Case [CL] with $\Lambda>0$ and Case
[RP] with $p>1$. In general, the similar behavior of the scale factor
is found in the case with $V>0$ and $\Sigma>0$.

Next, we consider Case [CF], with $f=1$, $D=4$, $q=2$, $t_0=0$
and $P=2$.
The similar model with the flux and the dilaton field has been studied
in the literature including Ref.~\cite{CGG,Roy}.
We show the values $A(t)=-S^{D-2}(t)\frac{d^2S^{2-D}(t)}{dt^2}$
versus $t$ for $\theta=-0.3\sim 0.5$ in FIG.~\ref{fig1}~(b) in Case [CF].
The acceleration is strengthened for finite values of $|\theta|$,
conversely to Case [CC].
In general, the similar behavior of the scale factor
is found in the cases with $V>0$ and $\Sigma<0$.

To summarize, in the model described by the action with an exponential
scalar potential (\ref{exp}) with $V>0$,
the noncommutativity reduces the cosmic acceleration if
$\alpha^2<\frac{D-1}{2(D-2)}$ while the noncommutativity enhances the 
acceleration if $\alpha^2>\frac{D-1}{2(D-2)}$.

%%%%%%%%%%%%%%%%%%%%%%%%%%%%%%%%%%%%%%%%%%%%%%%%%%%%%%%%%%%%%%%%%%%%%%%%%%%
%%%%%%%%%%%%%%%%%%%%%%%%%%%%%%%%%%%%%%%%%%%%%%%%%%%%%%%%%%%%%%%%%%%%%%%%%%%
%%%%%%%%%%%%%%%%%%%%%%%%%%%%%%%%%%%%%%%%%%%%%%%%%%%%%%%%%%%%%%%%%%%%%%%%%%%
\section{The quantum system}
\label{quant}
%%%%%%%%%%%%%%%%%%%%%%%%%%%%%%%%%%%%%%%%%%%%%%%%%%%%%%%%%%%%%%%%%%%%%%%%%%%
%%%%%%%%%%%%%%%%%%%%%%%%%%%%%%%%%%%%%%%%%%%%%%%%%%%%%%%%%%%%%%%%%%%%%%%%%%%
%%%%%%%%%%%%%%%%%%%%%%%%%%%%%%%%%%%%%%%%%%%%%%%%%%%%%%%%%%%%%%%%%%%%%%%%%%%

%%%%%%%%%%%%%%%%%%%%%%%%%%%%%%%%%%%%%%%%%%%%%%%%%%%%%%%%%%%%%%%%%%%%%%%%%%%
%%%%%%%%%%%%%%%%%%%%%%%%%%%%%%%%%%%%%%%%%%%%%%%%%%%%%%%%%%%%%%%%%%%%%%%%%%%
\subsection{Wave function of the universe}
\label{swf}
%%%%%%%%%%%%%%%%%%%%%%%%%%%%%%%%%%%%%%%%%%%%%%%%%%%%%%%%%%%%%%%%%%%%%%%%%%%
%%%%%%%%%%%%%%%%%%%%%%%%%%%%%%%%%%%%%%%%%%%%%%%%%%%%%%%%%%%%%%%%%%%%%%%%%%%

In a commutative model,
we can obtain the
minisuperspace Wheeler--De\,Witt equation by replacing
$\pi_x\rightarrow-i\frac{\partial}{\partial x}$ and
$\pi_y\rightarrow-i\frac{\partial}{\partial y}$ in the Hamiltonian $H$ and
regarding the Hamiltonian constraint as $H\Psi(x,y)=0$, where $\Psi(x,y)$ is the
wave function of the universe \cite{Halliwell}.

Deformation of the Wheeler-De~Witt equation in our noncommtative case
with the Hamiltonian $H_\theta$
can be performed by
\begin{eqnarray}
& &X\rightarrow\hat{X}=x+i\frac{\theta-\rho}{2}\frac{\partial}{\partial
y}\,,\quad Y\rightarrow\hat{Y}=y-i\frac{\theta+\rho}{2}\frac{\partial}{\partial
x}\,,\nonumber \\
& &\Pi_X\rightarrow\hat{\Pi}_X=-i\frac{\partial}{\partial x}
\,,\quad \Pi_Y\rightarrow\hat{\Pi}_Y=-i\frac{\partial}{\partial y}\,.
\end{eqnarray}
These operators satisfy the relations
\begin{equation}
[\hat{X}, \hat{\Pi}_X]=i\,,\quad
[\hat{Y}, \hat{\Pi}_Y]=i\,,\quad
[\hat{X}, \hat{Y}]=i\theta\,,
\end{equation}
and other commutators vanish.

The corresponding Wheeler-De~Witt equation is found to be
\begin{equation}
\frac{1}{2}\left\{\frac{\partial^2}{\partial x^2}-\frac{\partial^2}{\partial
y^2}+U \exp\left[2\lambda\left(x+i\frac{\theta-\rho}{2}\frac{\partial}{\partial
y}\right)\right]\right\}\Psi(x,y)=0\,.
\end{equation}
The solution of the equation can be written as \cite{GOR,ALNW}
\begin{equation}
\Psi(x, y)=\int_{-\infty}^\infty d\nu\, {\cal
A}_\nu\,\Psi_\nu(x,y)=\int_{-\infty}^\infty d\nu\, {\cal A}_\nu\,\psi_\nu(x)\,
e^{i\nu(y-y_0)}\,,
\label{wf}
\end{equation}
where
\begin{equation}
\psi_\nu(x)=c_1F_{i\nu/\lambda}(\sqrt{U}e^{\lambda
[x-\nu(\theta-\rho)/2]}/\lambda)+ c_2G_{i\nu/\lambda}(\sqrt{U}e^{\lambda
[x-\nu(\theta-\rho)/2]}/\lambda)\,,
\quad (U>0)
\label{Up}
\end{equation}
with
\begin{equation}
F_{\nu}(z)\equiv\frac{1}{2 \cos(\nu\pi/2)}[J_\nu(z)+J_{-\nu}(z)]\,,\quad
G_{\nu}(z)\equiv\frac{1}{2 \sin(\nu\pi/2)}[J_\nu(z)-J_{-\nu}(z)]
\,,
\end{equation}
and
\begin{equation}
\psi_\nu(x)=c_3K_{i\nu/\lambda}(\sqrt{|U|}e^{\lambda
[x-\nu(\theta-\rho)/2]}/\lambda)\,.\quad (U<0)
\label{Um}
\end{equation}
In the above expressions, ${\cal A}_\nu$ is the overall amplitude for $\nu$
and $c_1, c_2$ and $c_3$ are normalization factors.
$J_\nu(z)$ and $K_\nu(z)$ represent the Bessel function and the modified Bessel
function of the second kind, respectively.

In the noncommutative case,
we would like to consider $\Psi(X, Y)$ instead of $\Psi(x, y)$,
i.e., the expression in the main noncommutative variables.
%RR
In our model, the relation $\hat{\Pi}_Y\Psi_\nu=\nu\Psi_\nu$ holds.
Moreover, if we take the parameter choice $\rho=-\theta$, we find that $Y$ equals
$y$.
%RR
Therefore, we can regard $X\sim x-\theta P$ approximately at this parameter choice,
where
$P$ is the most probable value
%RR
of $\nu$
%RR
 for the peaked wave packet.

The Gaussian wave packet 
% setting the amplitude as
%${\cal A}_\nu=\frac{1}{\sqrt{\Gamma\sqrt{\pi}}}\exp\left[
%-\frac{(\nu-P)^2}{2\Gamma^2}\right]$
has often been considered  many
times \cite{ALNW,KN,Kiefer1,Kiefer2} in a semiclassical analysis of
quantum cosmology.
For simplicity in calculations, we adopt  here a rectangular form of the amplitude,
which is finite in the width $\Gamma$ around the central value $\nu=P$,
i.e.,
\begin{equation}
{\cal A}_\nu=\Gamma^{-1/2}\quad (P-\Gamma/2<\nu<P+\Gamma/2)\,,\mbox{~and~}
{\cal A}_\nu=0\quad  (\nu<P-\Gamma/2,\, P+\Gamma/2<\nu)\,,
\end{equation}
where $\Gamma$ is a constant.
Fortunately, we will find that the dependence on noncommutativity has more
influence than the choice of the shape of amplitude.

First, we consider the case with $U>0$ (\ref{Up}).
We set the factors as \cite{ALNW}
\begin{equation}
c_1=\sqrt{\frac{\nu/\lambda}{\tanh(\nu\pi/(2\lambda))}}\,,\quad
c_2=0\,.
\end{equation}
The choice that $c_1=0$ and $c_2\ne 0$ and the other general choices 
will lead to a similar result, because $F_\nu(e^x)$ and $G_\nu(e^x)$ have only
the different phase shift in $x$ \cite{ALNW}.

\begin{figure}[ht]
\centering
\includegraphics[width=5cm]{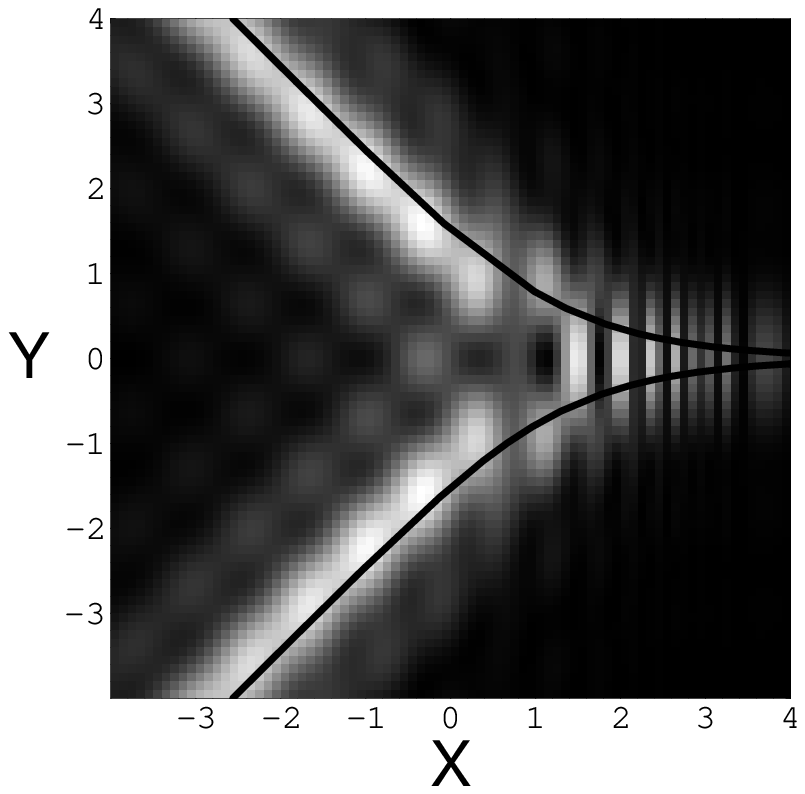}
\includegraphics[width=5cm]{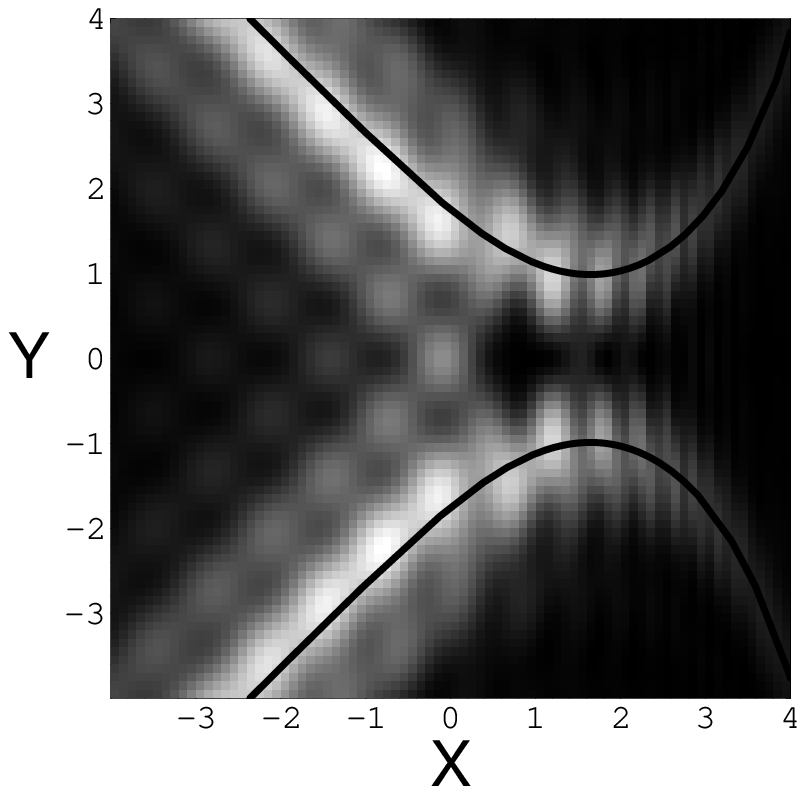}
\includegraphics[width=5cm]{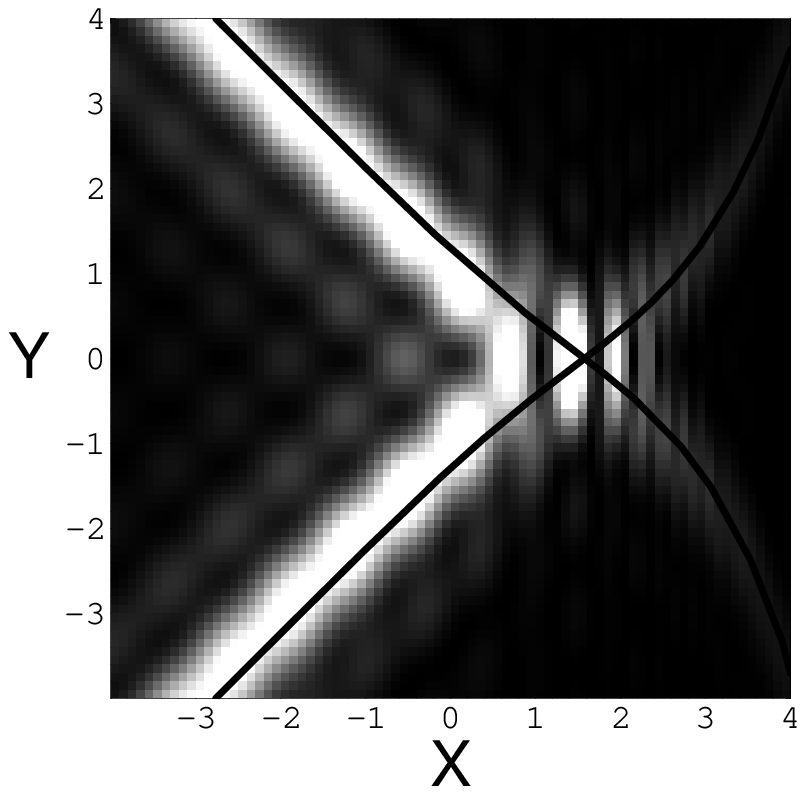}
\\
\hspace{0.5cm} (a) \hspace{4.5cm} (b) \hspace{4.5cm} (c)
\caption{The probability
distribution $\varrho\equiv|\Psi(X,Y)|^2$ in the case with $U>0$. The figures
correspond to the case with: (a) $\theta=0$, (b) $\theta=0.1$, (c)
$\theta=-0.1$. 
For the other parameters, see text.}
\label{fig2}
\end{figure}

We show the density plot of the probability distribution
$\varrho(X,Y)\equiv|\Psi(X,Y)|^2$ with the parameter
$\rho=-\theta$ against
$X\equiv x-\theta P$ and
$Y=y$ in the case with $U>0$ in FIG.~\ref{fig2}.
Here, we consider Case [CC] with $k_b=-1$, $D=4$, $q=2$, $t_0=0$
and $P=2$ and set $\Gamma=5$.
The figures correspond to the case with: (a) $\theta=0$, (b) $\theta=0.1$, (c)
$\theta=-0.1$. 
The solid curves in figures shows the classical solutions (\ref{csp}).

%%%%%%%%%%%%%%%%%%%%%%%%%%%
% 2
%%%%%%%%%%%%%%%%%%%%%%%%%%%
%\begin{wrapfigure}{r}{5cm}
%\end{wrapfigure}
%%%%%%%%%%%%%%%%%%%%%%%%%%%

Next, we consider the case with $U<0$ (\ref{Um}).
We set the factors as 
\begin{equation}
c_3=\sqrt{\frac{2(\nu/\lambda)\sinh(\pi\nu/\lambda)}{\pi}}\,.
\end{equation}

\begin{figure}[ht]
\centering
\includegraphics[width=5cm]{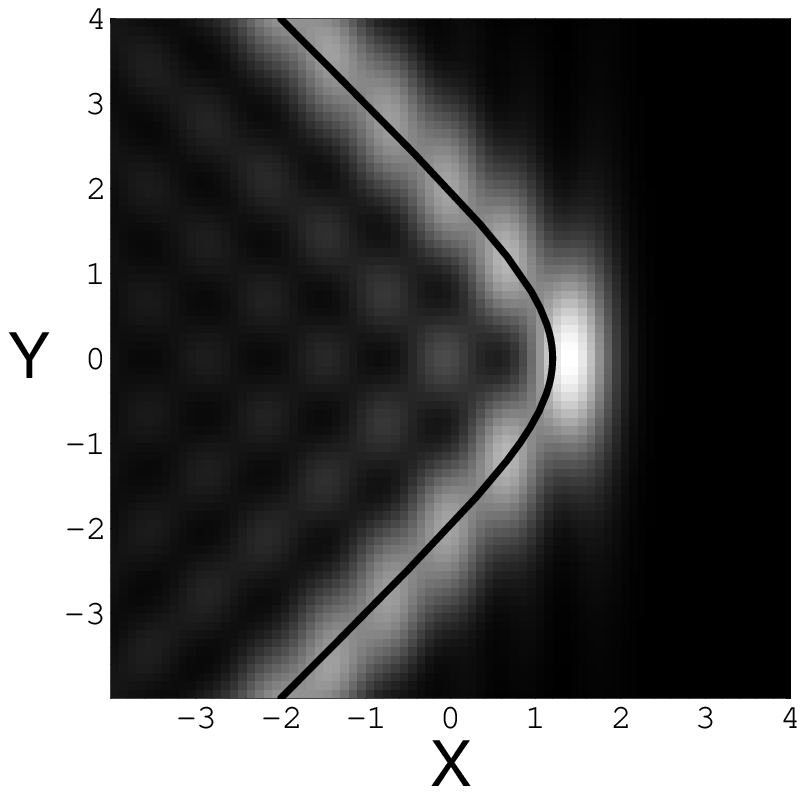}
\includegraphics[width=5cm]{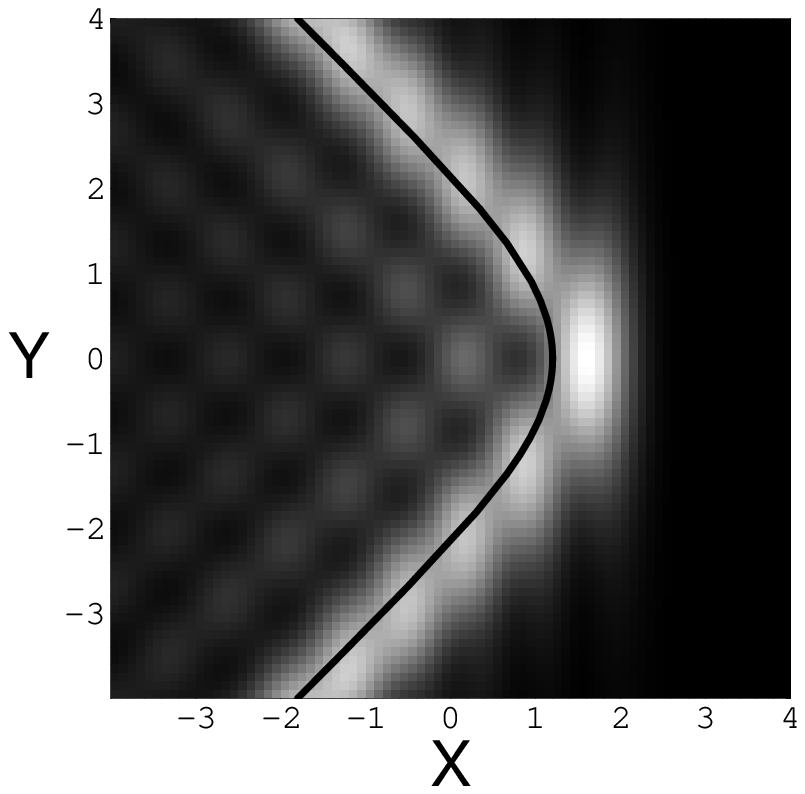}
\includegraphics[width=5cm]{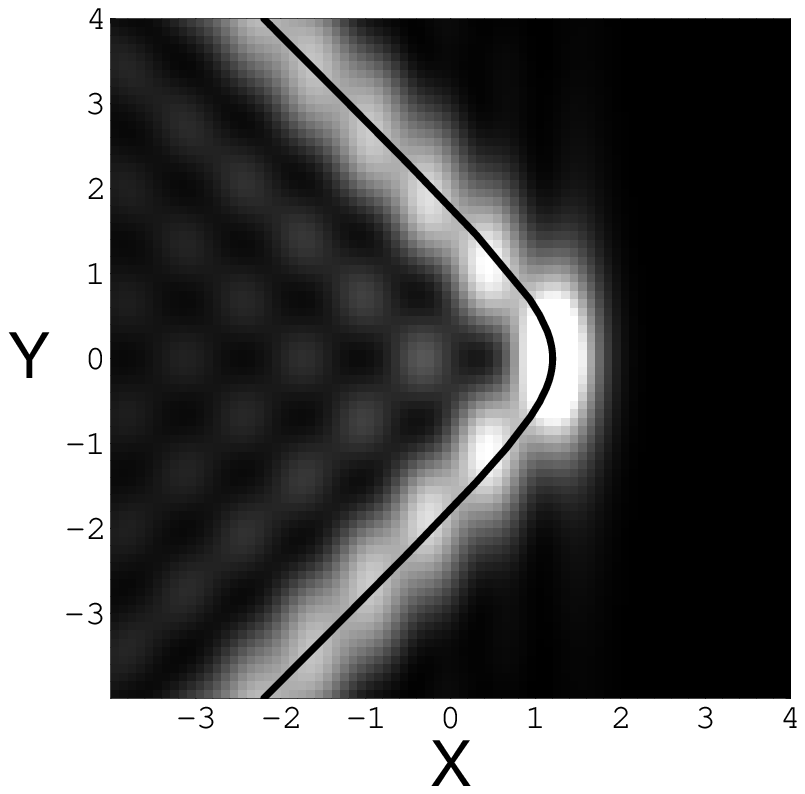}
\\
\hspace{0.5cm} (a) \hspace{4.5cm} (b) \hspace{4.5cm} (c)
\caption{The probability distribution $\varrho\equiv|\Psi(X,Y)|^2$
in the case with $U<0$. The figures correspond to the case with: (a) $\theta=0$,
(b) $\theta=0.1$, (c) $\theta=-0.1$. 
For the other parameters, see text.}
\label{fig3}
\end{figure}

We plot  $\varrho=|\Psi(X,Y)|^2$ with the
parameter
$\rho=-\theta$ against
$X=x-\theta P$ and
$Y=y$ in the case with $U<0$ in FIG.~\ref{fig3}.
Here, we consider Case [CF] with $f=1$, $D=4$, $q=2$, $t_0=0$
and $P=2$ and set $\Gamma=5$.
The figures correspond to the case with: (a) $\theta=0$, (b) $\theta=0.1$, (c)
$\theta=-0.1$. The solid curves in figures shows the classical solutions
(\ref{csm}).

%%%%%%%%%%%%%%%%%%%%%%%%%%%
% 3
%%%%%%%%%%%%%%%%%%%%%%%%%%%
%\begin{wrapfigure}{r}{5cm}
%\end{wrapfigure}
%%%%%%%%%%%%%%%%%%%%%%%%%%%

In each case, one finds that the peak of the distribution function reproduces
a classical trajectory well, even with the simplest rectangular form of
amplitude adopted here. 

The interpretation of the noncommutative variables,
especially $X$ in the present case
may not be accepted in general noncommutative dynamics,
since the `expectation value' $P$ for $\nu$ corresponds to a unique constant in our
classical model.

%%%%%%%%%%%%%%%%%%%%%%%%%%%%%%%%%%%%%%%%%%%%%%%%%%%%%%%%%%%%%%%%%%%%%%%%%%%
%%%%%%%%%%%%%%%%%%%%%%%%%%%%%%%%%%%%%%%%%%%%%%%%%%%%%%%%%%%%%%%%%%%%%%%%%%%
\subsection{Wigner function and deformation}
\label{wigner}
%%%%%%%%%%%%%%%%%%%%%%%%%%%%%%%%%%%%%%%%%%%%%%%%%%%%%%%%%%%%%%%%%%%%%%%%%%%
%%%%%%%%%%%%%%%%%%%%%%%%%%%%%%%%%%%%%%%%%%%%%%%%%%%%%%%%%%%%%%%%%%%%%%%%%%%
%v2
In the previous subsection, we utilized the expectation value of $p_y\approx\nu$ to
define the noncommutative variable $X$. In general, the noncommutative variable
can be obtained from the linear combination of phase space variables, say, $x$ and
$p_y$ in the two dimensional commutative system.
 Now,
we propose another candidate of probability distribution by using the Wigner
function \cite{Wigner,WF} and examine its validity by analysis with our soluble
model. The Wigner function gives a (pseudo) probability distribution whose
arguments are dynamical variables and their conjugate variables.
%v2

Generally speaking, the Wigner function \cite{Wigner,WF} is defined, in terms of a
wave function
$\phi(q)$, by
\begin{equation}
W(q, p)\equiv \frac{1}{2\pi}\int_{-\infty}^\infty du\,
\phi^*\left(q-\frac{u}{2}\right)\phi\left(q+\frac{u}{2}\right)e^{-ipu}\,.
\end{equation}
The Wigner function has beautiful properties, such as
\begin{equation}
\int_{-\infty}^\infty dp\,W(q,
p)=\left|\phi\left(q\right)\right|^2\,,\quad
\int_{-\infty}^\infty dq\,W(q,
p)=|\tilde{\phi}\left(p\right)|^2\,,
\end{equation}
where $\tilde{\phi}(p)$ is the Fourier transform of $\phi(q)$.

The application of the Wigner function has been considered 
already in quantum cosmologies with deformed phase spaces with slightly
different motivations from ours
\cite{CGT,RJ}.  Our present aim is to construct the probability distribution whose
arguments are noncommutative variables, say, $X$ and $Y$, from the Wigner function
defined by the wave function of commutative variables $x$ and $y$.

We start with the Wigner distribution function constructed from the wave function
$\Psi(x, y)$ (\ref{wf}), where we leave the variable $x$
untouched:
\begin{eqnarray}
W(x, y, p_y)&=&\frac{1}{2\pi}\int_{-\infty}^\infty dv\,
\Psi^*\left(x, y-\frac{v}{2}\right)\Psi\left(x,
y+\frac{v}{2}\right)e^{-ip_yv}\nonumber \\
&=&\frac{1}{2\pi}
\int_{-\infty}^\infty d\nu'\int_{-\infty}^\infty d\nu\,
{\cal A}_{\nu'}^*\psi^*_{\nu'}(x){\cal A}_{\nu}\psi_{\nu}(x)
e^{i(\nu-\nu')(y-y_0)}\int_{-\infty}^\infty dv\,e^{-i[p_y-(\nu'+\nu)/2]v}
\nonumber \\
&=&
\int_{-\infty}^\infty d\nu\,
{\cal A}_{p_y-\nu/2}^*\psi^*_{p_y-\nu/2}(x){\cal A}_{p_y+\nu/2}\psi_{p_y+\nu/2}(x)
e^{i\nu(y-y_0)}\,.
\end{eqnarray}
Therefore, the Fourier transform between $y$ and $\nu$ is easily obtained as a
simple form:
\begin{eqnarray}
\tilde{W}(x,\nu,p_y)&\equiv&
 \frac{1}{2\pi}\int_{-\infty}^\infty dy\,W(x,y,p_y)e^{-i\nu(y-y_0)}\nonumber \\
&=&{\cal A}_{p_y-\nu/2}^*\psi^*_{p_y-\nu/2}(x){\cal
A}_{p_y+\nu/2}\psi_{p_y+\nu/2}(x)\,.
\end{eqnarray}
Now, we come to an idea that one may interpret $X=x-\theta p_y$ as in the relation
(\ref{xX}) with $\rho=-\theta$.
To this end, we define
\begin{equation}
X\equiv x-\theta p_y\,,\quad \tilde{X}\equiv \theta x+p_y\,,
\end{equation}
and integrate out the extra variable $\tilde{X}$.
Namely, we propose a (Fourier transformed) probability distribution
\begin{equation}
\tilde{\varrho}_W(X, \nu)\equiv\frac{1}{1+\theta^2}
\int_{-\infty}^\infty d\tilde{X}\,\tilde{W}(x(X,\tilde{X}),
\nu, p_y(X,\tilde{X}))\,,
\end{equation}
where
\begin{equation}
x(X,\tilde{X})=\frac{1}{1+\theta^2} (X+\theta\tilde{X})\,,\quad 
p_y(X,\tilde{X})=\frac{1}{1+\theta^2} (\tilde{X}-\theta X)\,.
\end{equation}

To compare the new probability distribution with that considered in the last
section, we consider the Fourier transform of $\rho(X,Y)$:
\begin{eqnarray}
\tilde{\cal \varrho}(X,\nu)
&=&\frac{1}{2\pi}\int_{-\infty}^\infty dy
\int_{-\infty}^\infty d\nu''\int_{-\infty}^\infty d\nu'\,
{\cal A}_{\nu''}^*\psi^*_{\nu''}(X+\theta P){\cal A}_{\nu'}\psi_{\nu'}(X+\theta P)
e^{i(\nu'-\nu'')(y-y_0)}\,e^{-i\nu(y-y_0)}
\nonumber \\
&=&
\int_{-\infty}^\infty d\nu'\,
{\cal A}_{\nu'-\nu/2}^*\psi^*_{\nu'-\nu/2}(X+\theta P){\cal
A}_{\nu'+\nu/2}\psi_{\nu'+\nu/2}(X+\theta P)\,.
\end{eqnarray}
These expressions show the fact that $\tilde{\cal \varrho}_W(X,\nu)$ can be very
close to $\tilde{\cal \varrho}(X,\nu)$ if the amplitude ${\cal A}_\nu$
has the compact form of which central value is located at $\nu\sim P$.
One can also find that $\tilde{\cal \varrho}_W(X,\nu)$=$\tilde{\cal
\varrho}(X,\nu)$ exactly in the commutative case with $\theta=0$.

%%%%%%%%%%%%%%%%%%%%%%%%%%%
% 4
%%%%%%%%%%%%%%%%%%%%%%%%%%%
%\begin{wrapfigure}{r}{5cm}
\begin{figure}[ht]
\centering
\includegraphics[width=5cm]{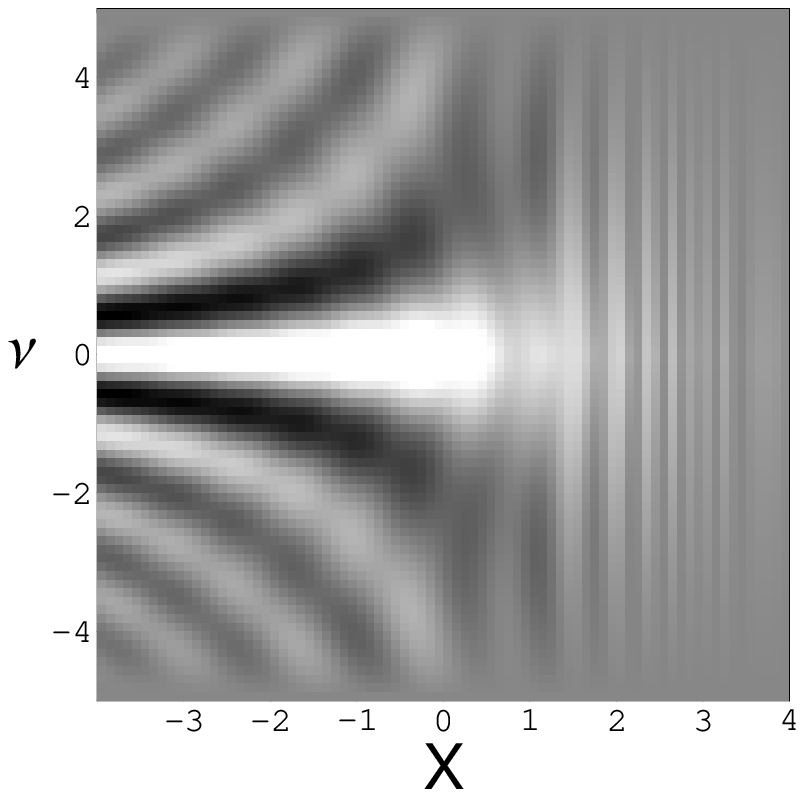}
\includegraphics[width=5cm]{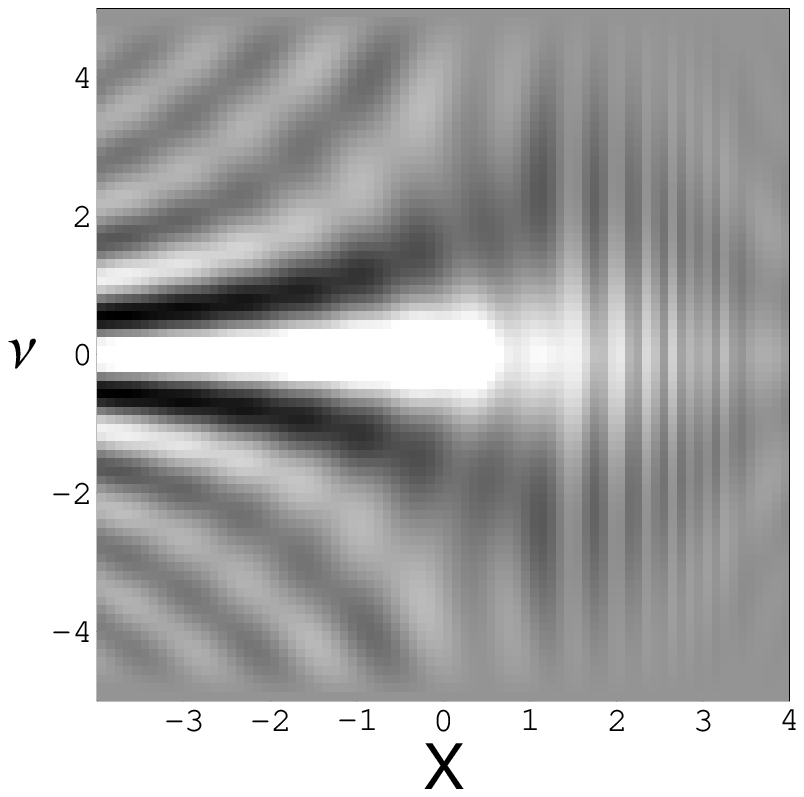}
\includegraphics[width=5cm]{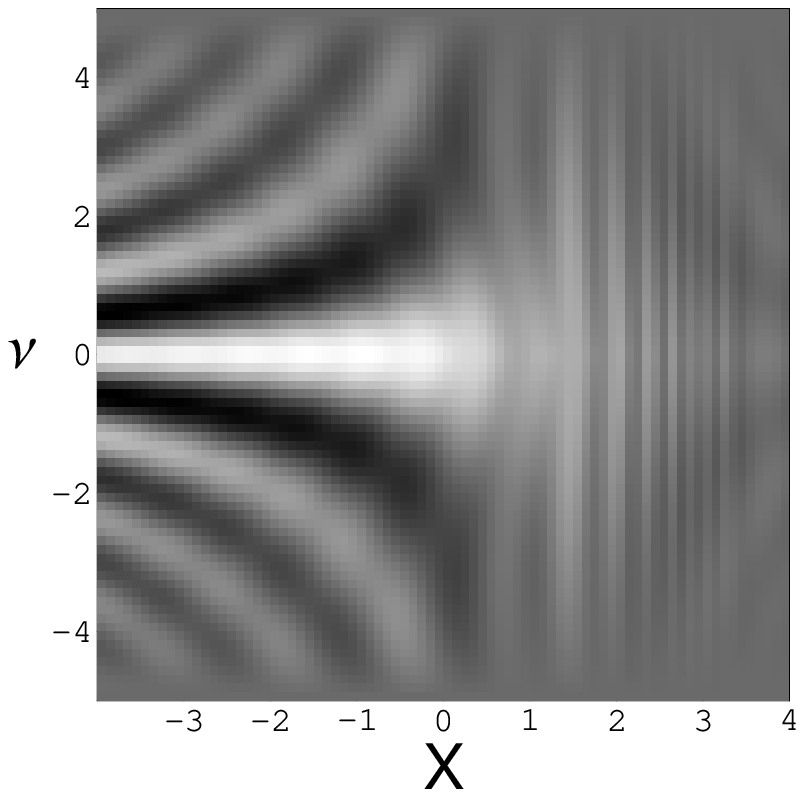}
\\
\hspace{0.5cm} (a) \hspace{4.5cm} (b) \hspace{4.5cm} (c)
\caption{The new Fourier transformed probability distribution
$\tilde{\varrho}_W(X,\nu)$ in the case with $U>0$. The figures correspond to
the case with: (a) $\theta=0$, (b) $\theta=0.1$, (c) $\theta=-0.1$.}
\label{fig4}
\end{figure}
%\end{wrapfigure}
%%%%%%%%%%%%%%%%%%%%%%%%%%%

%%%%%%%%%%%%%%%%%%%%%%%%%%%
% 5
%%%%%%%%%%%%%%%%%%%%%%%%%%%
%\begin{wrapfigure}{r}{5cm}
\begin{figure}[ht]
\centering
\includegraphics[width=5cm]{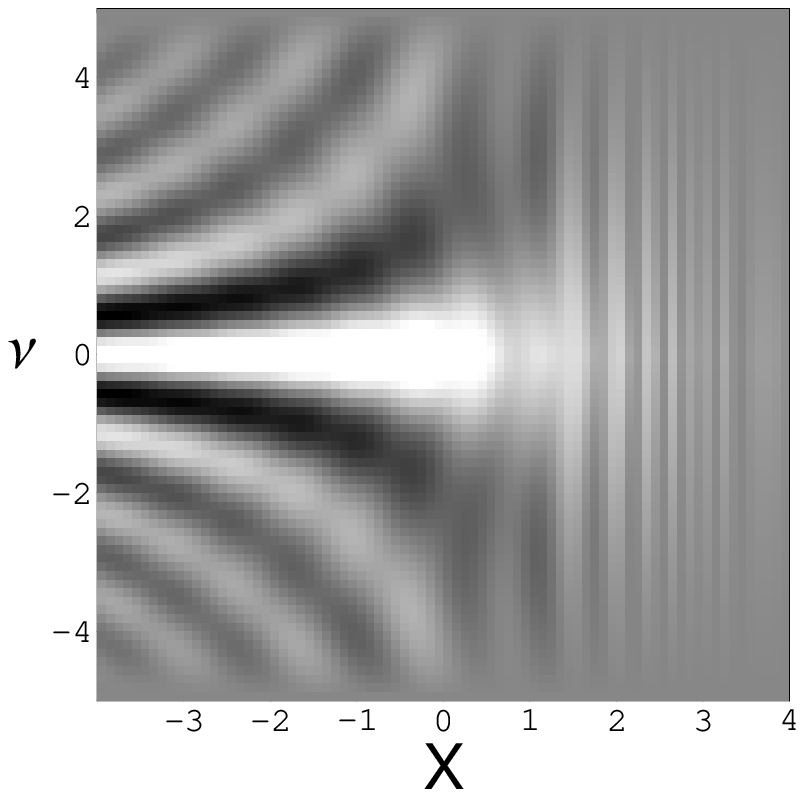}
\includegraphics[width=5cm]{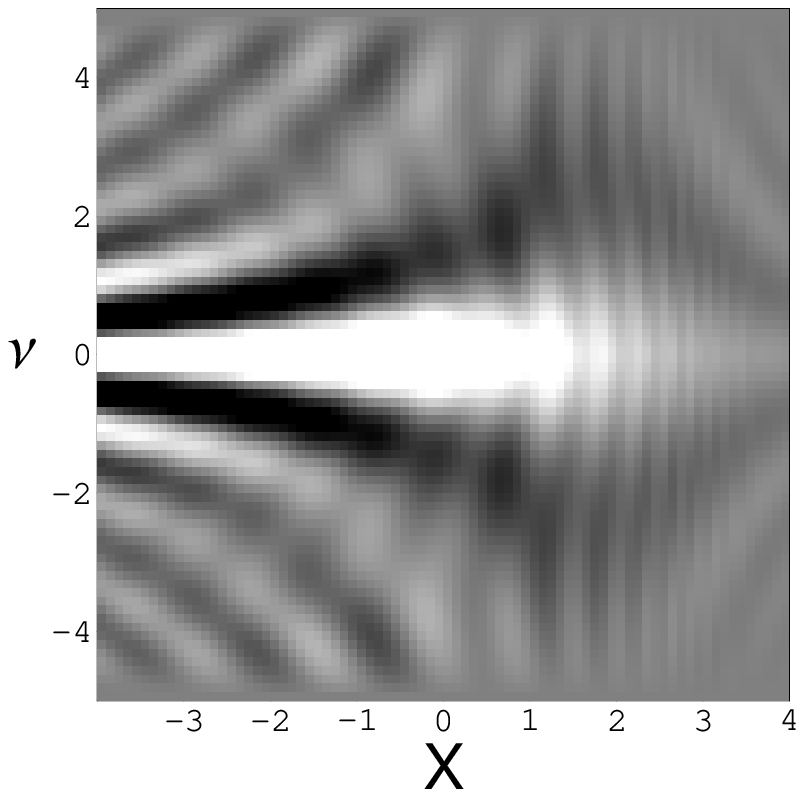}
\includegraphics[width=5cm]{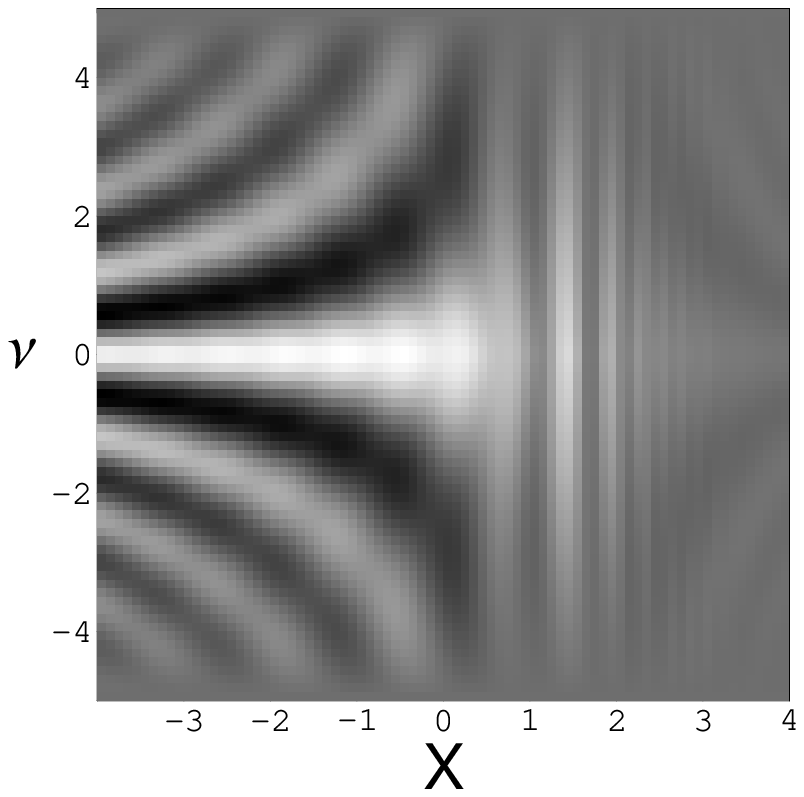}
\\
\hspace{0.5cm} (a) \hspace{4.5cm} (b) \hspace{4.5cm} (c)
\caption{The Fourier transform of the probability distribution
defined in the previous section,
$\tilde{\varrho}(X,\nu)$ in the case with $U>0$ for comparison. The figures
correspond to the case with: (a) $\theta=0$, (b) $\theta=0.1$, (c) $\theta=-0.1$.}
\label{fig5}
\end{figure}
%\end{wrapfigure}
%%%%%%%%%%%%%%%%%%%%%%%%%%%

%%%%%%%%%%%%%%%%%%%%%%%%%%%
% 6
%%%%%%%%%%%%%%%%%%%%%%%%%%%
%\begin{wrapfigure}{r}{5cm}
\begin{figure}[ht]
\centering
\includegraphics[width=5cm]{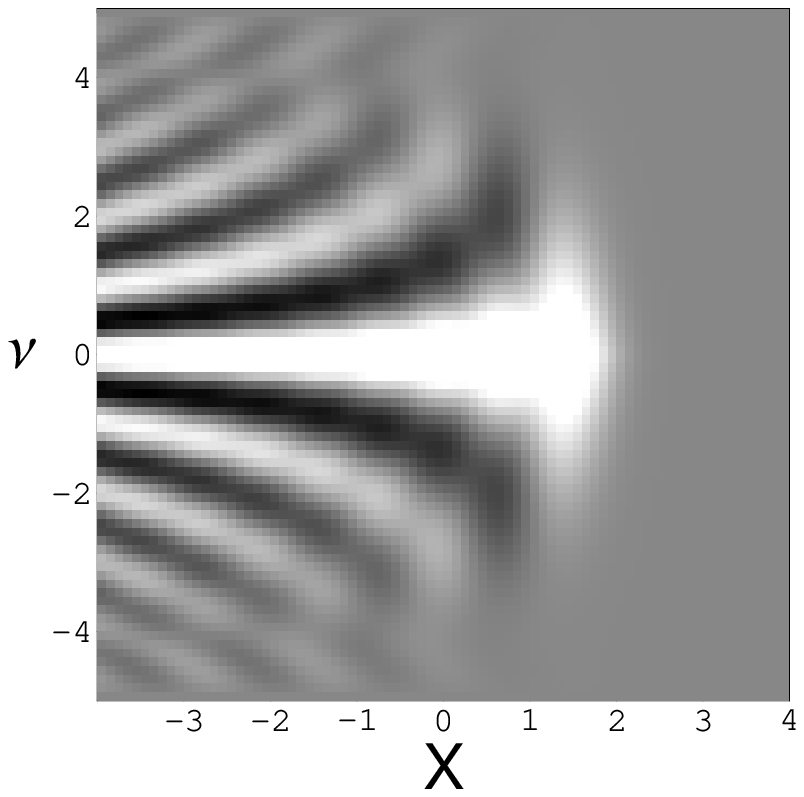}
\includegraphics[width=5cm]{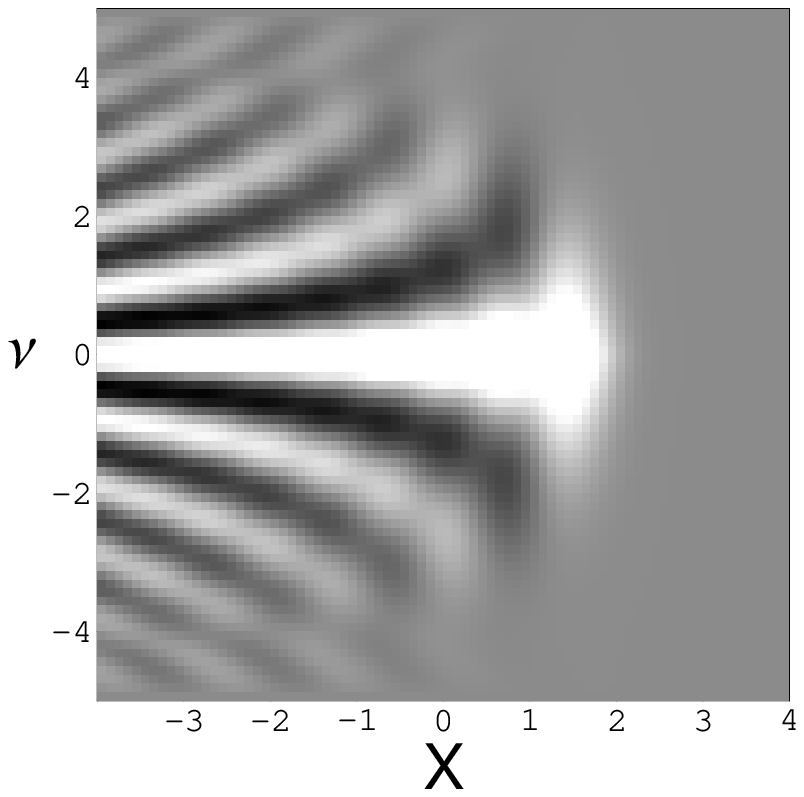}
\includegraphics[width=5cm]{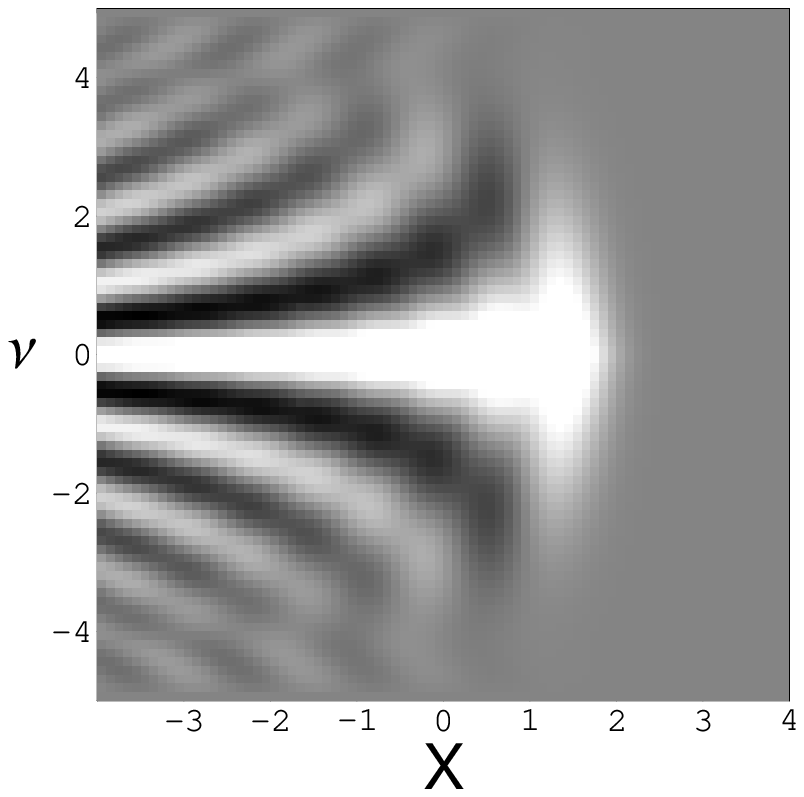}
\\
\hspace{0.5cm} (a) \hspace{4.5cm} (b) \hspace{4.5cm} (c)
\caption{The new Fourier transformed probability distribution
$\tilde{\varrho}_W(X,\nu)$ in the case with $U<0$. The figures correspond to
the case with: (a) $\theta=0$, (b) $\theta=0.1$, (c) $\theta=-0.1$.}
\label{fig6}
\end{figure}
%\end{wrapfigure}
%%%%%%%%%%%%%%%%%%%%%%%%%%%

%%%%%%%%%%%%%%%%%%%%%%%%%%%
% 7
%%%%%%%%%%%%%%%%%%%%%%%%%%%
%\begin{wrapfigure}{r}{5cm}
\begin{figure}[ht]
\centering
\includegraphics[width=5cm]{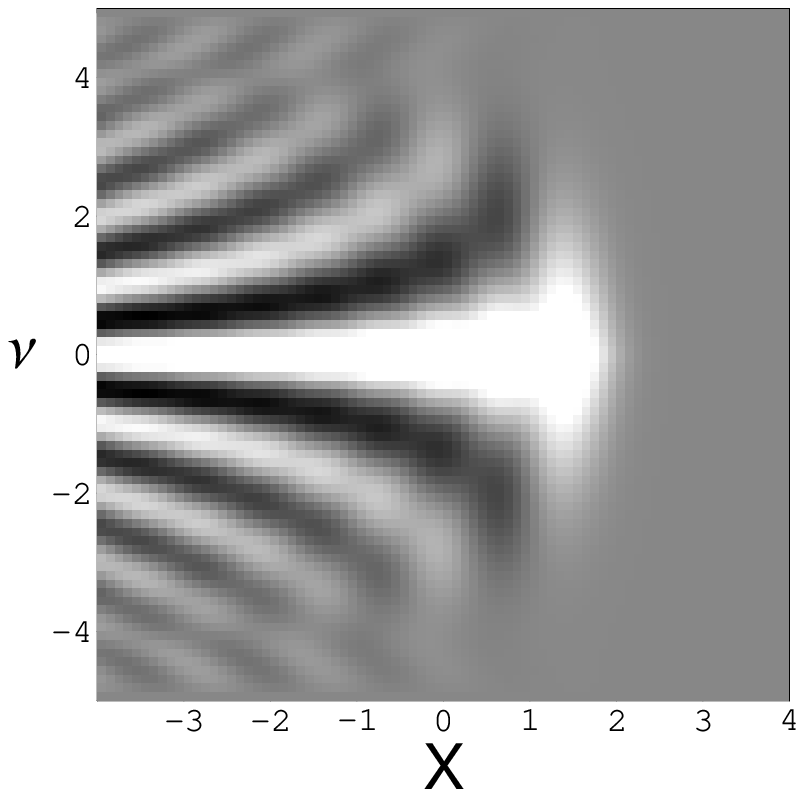}
\includegraphics[width=5cm]{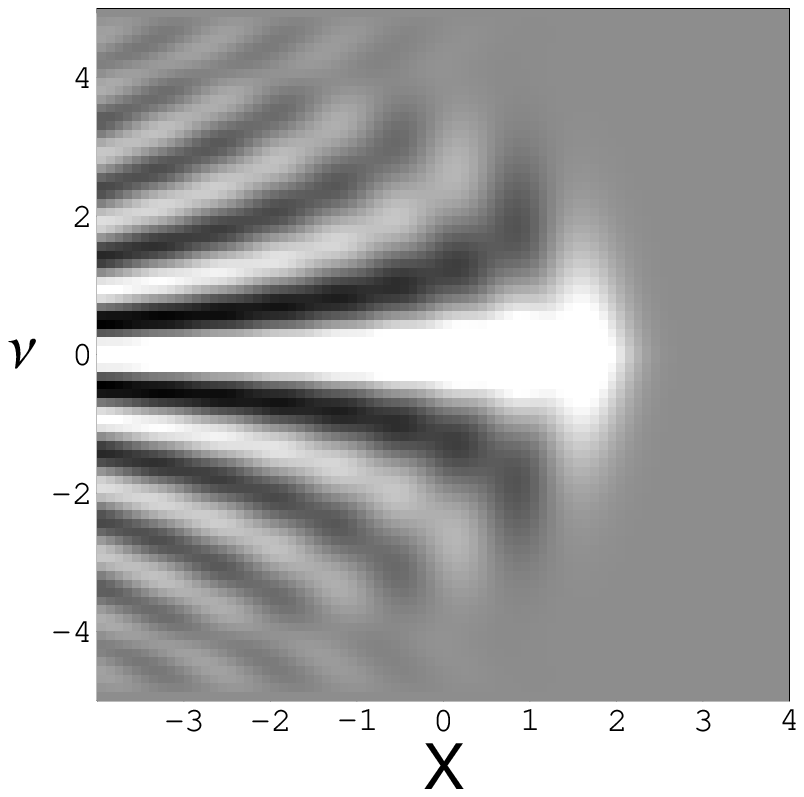}
\includegraphics[width=5cm]{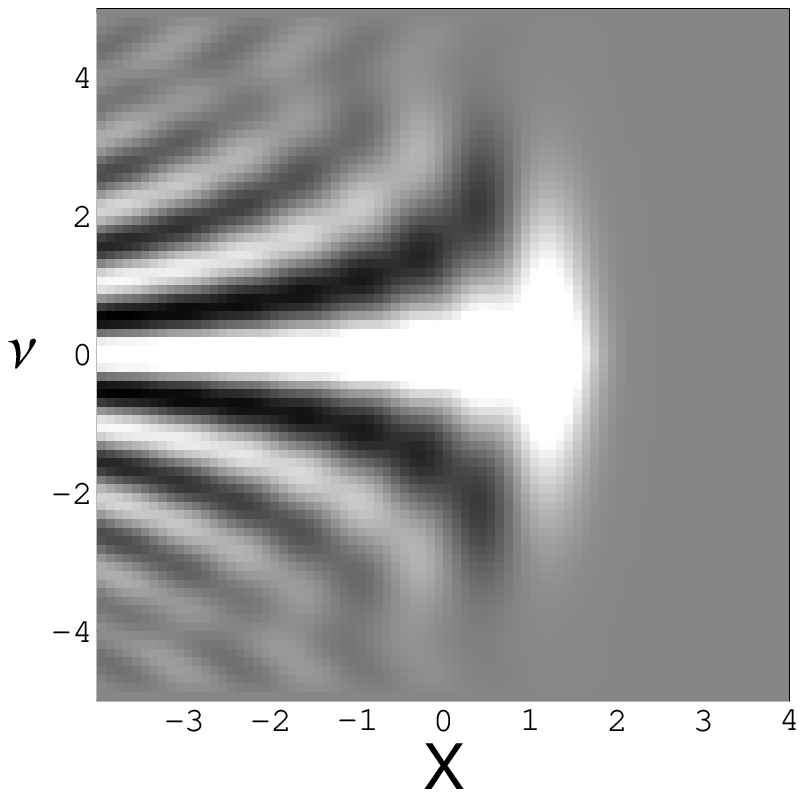}
\\
\hspace{0.5cm} (a) \hspace{4.5cm} (b) \hspace{4.5cm} (c)
\caption{The Fourier transform of the probability distribution
defined in the previous section,
$\tilde{\varrho}(X,\nu)$ in the case with $U<0$ for comparison. The figures
correspond to the case with: (a) $\theta=0$, (b) $\theta=0.1$, (c) $\theta=-0.1$.}
\label{fig7}
\end{figure}
%\end{wrapfigure}
%%%%%%%%%%%%%%%%%%%%%%%%%%%

The numerical calculations for the probability distributions are shown in
FIG.~\ref{fig4}, FIG.~\ref{fig5}, FIG.~\ref{fig6}, and FIG.~\ref{fig7}. 
The case with $U>0$ is shown in FIG.~\ref{fig4} and FIG.~\ref{fig5}.
FIG.~\ref{fig4} represents $\tilde{\varrho}_W(X,Y)$ while FIG.~\ref{fig5}
represents $\tilde{\varrho}(X,Y)$ which is the Fourier transform of the
distribution $\varrho(X,Y)$ considered in the previous section.
In the figures, the noncommutative parameter is set to: (a) $\theta=0$,
(b) $\theta=0.1$, and (c) $\theta=-0.1$.
One can hardly find notable differences between FIG.~\ref{fig4} and
FIG.~\ref{fig5}. Similarly, the case with $U<0$ is shown in FIG.~\ref{fig6} and
FIG.~\ref{fig7}. Any remarkable difference cannot be found also in this case.

Consequently, it is safely to say that the newly defined distribution function
$\tilde{\varrho}_W(X,Y)$ is valid for describing 
the present model. 
Note that our present model and analysis may be very specific;
we have only studied quantum cosmologies in terms of wave packets, which have
rather semiclassical properties and is easy to treat.
The study of the probability
distribution derived from the Wigner function in terms of generic wave functions
will be expected in future work. Nonetheless, the verification of the validity
through the exact analytic solutions in the present work is the first essential
step to study the noncommutative quantum cosmology with noncommutative variables.

%%%%%%%%%%%%%%%%%%%%%%%%%%%%%%%%%%%%%%%%%%%%%%%%%%%%%%%%%%%%%%%%%%%%%%%%%%%
%%%%%%%%%%%%%%%%%%%%%%%%%%%%%%%%%%%%%%%%%%%%%%%%%%%%%%%%%%%%%%%%%%%%%%%%%%%
%%%%%%%%%%%%%%%%%%%%%%%%%%%%%%%%%%%%%%%%%%%%%%%%%%%%%%%%%%%%%%%%%%%%%%%%%%%
\section{discussion and outlook}
\label{do}
%%%%%%%%%%%%%%%%%%%%%%%%%%%%%%%%%%%%%%%%%%%%%%%%%%%%%%%%%%%%%%%%%%%%%%%%%%%
%%%%%%%%%%%%%%%%%%%%%%%%%%%%%%%%%%%%%%%%%%%%%%%%%%%%%%%%%%%%%%%%%%%%%%%%%%%
%%%%%%%%%%%%%%%%%%%%%%%%%%%%%%%%%%%%%%%%%%%%%%%%%%%%%%%%%%%%%%%%%%%%%%%%%%%

In this paper, a noncommutative deformation of the minisuperspace
variables is studied by means of an integrable model. Its analytical solutions
are obtained in classical and quantum cosmology.

It has been already known that the $D$-dimensional model with an exponential scalar
potential
$\frac{V}{2} e^{2\alpha\phi}$, which is related to
higher-dimensional/higher-derivative theories, gives rise to an accelerating
universe. We find that the noncommutativity suppresses acceleration if the
coupling $\alpha$ is small whereas it enhances acceleration if the coupling
$\alpha$ is large%
%RR
, provided that $V>0$.
%RR
The critical value is given by $\alpha^2={\frac{D-1}{2(D-2)}}$
in our model.

%v3
The different behaviors, which depend on the sign of $U$, should be thoroughly
investigated both in classical and quantum perspectives, since interesting
features in the Liouville quantum mechanics with a negative potential have been
reported \cite{KT}. This issue is left for a future work.
%v3

We have managed to interpret the probability distribution in noncommutative
quantum cosmology. We first showed that the peak of the wave packet
reproduces the classical trajectory by using exact solutions with an interpretation
of the noncommutative variables in the present model.
Next, we proposed a new probability distribution in noncommutative
quantum cosmology constructed from the Wigner function.
Its validity in the present solvable model is confirmed numerically.

%v3
In general, the Wigner function is not positive-definite, as is well-known. 
Although our study focusing on wave packets is not suffering from
the problem of `negative probability', we should treat carefully the problem with
general settings and we have to search for a possible solution to the problem.
%v3

In future study, we will investigate general noncommutative cosmology
by using the probability distribution function provided in this paper.
Also, general deformations of minisuperspace variables
 should be studied further.
The model with a phantom scalar field \cite{phantom,DKS,ALNW} and/or
a phantom gauge field \cite{KS} may also be worth studying in the context of
noncommutative cosmology.

%%%%%%%%%%%%%%%%%%%%%%%%%%%%%%%%%%%%%%%%%%%%%%%%%%%%%%%%%%%%%%%%%
%%%%%%%%%%%%%%%%%%%%%%%%%%%%%%%%%%%%%%%%%%%%%%%%%%%%%%%%%%%%%%%%%
%\appendix
%%%%%%%%%%%%%%%%%%%%%%%%%%%%%%%%%%%%%%%%%%%%%%%%%%%%%%%%%%%%%%%%%
%%%%%%%%%%%%%%%%%%%%%%%%%%%%%%%%%%%%%%%%%%%%%%%%%%%%%%%%%%%%%%%%%

%%%%%%%%%%%%%%%%%%%%%%%%%%%%%%%%%%%%%%%%%%%%%%%%%%%%%%%%%%%%%%%%%%%%%%%%%%%
%\acknowledgments
%%%%%%%%%%%%%%%%%%%%%%%%%%%%%%%%%%%%%%%%%%%%%%%%%%%%%%%%%%%%%%%%%%%%%%%%%%%
%Acknowledgements
%%%%%%%%%%%%%%%%%%%%%%%%%%%%%%%%%%%%%%%%%%%%%%%%%%%%%%%%%%%%%%%%%%%%%%%%%%%
%\begin{acknowledgments}
%We thank
%the organizers of JGRG21, where our
%partial result %({\tt [arXiv:10mm.xxxx]}) 
%was presented. %for elucidating comments.
%This study is supported in part by the Grant-in-Aid of Nikaido Research 
%Fund.
%\end{acknowledgments}
%%%%%%%%%%%%%%%%%%%%%%%%%%%%%%%%%%%%%%%%%%%%%%%%%%%%%%%%%%%%%%%%%%%%%%%%%%%

%%%%%%%%%%%%%%%%%%%%%%%%%%%%%%%%%%%%%%%%%
%%%%%%%%%%%%%%%%%%%%%%%%%%%%%%%%%%%%%%%%%
%%%
%%%   References
%%%
%%%%%%%%%%%%%%%%%%%%%%%%%%%%%%%%%%%%%%%%%
%%%%%%%%%%%%%%%%%%%%%%%%%%%%%%%%%%%%%%%%%
%%%%%%%%%%%%%%%%%%%%%%%%%%%%%%%%%%%%%%%%%%%%%%%%%%%%%%%%%%%%%%%%%%%%%%%%%%%
%thebibliography
%%%%%%%%%%%%%%%%%%%%%%%%%%%%%%%%%%%%%%%%%%%%%%%%%%%%%%%%%%%%%%%%%%%%%%%%%%%
%\bibliographystyle{apsrev}
\bibliographystyle{apsrev4-1}
%\bibliography{}

%%%%%%%%%%%%%%%%%%%%%%%%%%%%%%%%%%%%%%%%%%%%%%%%%%%%%%%%%%%%%%%%%%%%%%%%%%%
%%%%%%%%%%%%%%%%%%%%%%%%%%%%%%%%%%%%%%%%%%%%%%%%%%%%%%%%%%%%%%%%%%%%%%%%%%%
%%%%%%%%%%%%%%%%%%%%%%%%%%%%%%%%%%%%%%%%%%%%%%%%%%%%%%%%%%%%%%%%%%%%%%%%%%%
\end{document}